\documentclass[journal]{IEEEtran}

\usepackage[utf8]{inputenc}
\usepackage[T1]{fontenc}
\usepackage{amsmath}
\usepackage{cite}
\usepackage{xcolor}

\usepackage{amssymb}
\usepackage{blkarray}
\usepackage{xtab,afterpage}
\usepackage{subcaption}
\usepackage[pdftex]{graphicx}
\usepackage{threeparttable}

\ifCLASSINFOpdf
\else
\fi

\begin{document}
%



\title{Inferring Gene Regulatory Neural Networks for Bacterial Decision Making in Biofilms}

\author{Samitha~Somathilaka,~\IEEEmembership{Student,~IEEE,}
        Daniel~P.~Martins,~\IEEEmembership{Member,~IEEE,}
        Xu~Li,
        Yusong~Li, 
        Sasitharan~Balasubramaniam,~\IEEEmembership{Senior Member,~IEEE,}
\thanks{Samitha Somathilaka is with VistaMilk Research Centre, Walton Institute for Information and Communication Systems Science, Waterford Institute of Technology, Waterford, X91 P20H, Ireland and School of Computing, University of Nebraska-Lincoln, 104 Schorr Center, 1100 T Street, Lincoln, NE, 68588-0150, USA. E-mail: {samitha.somathilaka}@waltoninstitute.ie.}
\thanks{Daniel P. Martins are with VistaMilk Research Centre and the Walton Institute for Information and Communication Systems Science, Waterford Institute of Technology, Waterford, X91 P20H, Ireland. E-mail: {daniel.martins}@waltoninstitute.ie.}
\thanks{Xu Li and Yusong Li are with Department of Civil and Environmental Engineering University of Nebraska-Lincoln 900 N. 16th Street Nebraska Hall W181, Lincoln, NE 68588-0531 E-mail:{xuli,yli7}@unl.edu.}
\thanks{S. Balasubramaniam is with School of Computing, University of Nebraska-Lincoln, 104 Schorr Center, 1100 T Street, Lincoln, NE, 68588-0150, USA. E-mail:sasi@unl.edu}}

%
%


\maketitle

\begin{abstract}
Bacterial cells are sensitive to a range of external signals used to learn the environment. These incoming external signals are then processed using a Gene Regulatory Network (GRN), exhibiting similarities to modern computing algorithms. An in-depth analysis of gene expression dynamics suggests an inherited Gene Regulatory Neural Network (GRNN) behavior within the GRN that enables the cellular decision-making based on received signals from the environment and neighbor cells. In this study, we extract a sub-network of \textit{Pseudomonas aeruginosa} GRN that is associated with one virulence factor: pyocyanin production as a use case to investigate the GRNN behaviors. Further, using Graph Neural Network (GNN) architecture, we model a single species biofilm to reveal the role of GRNN dynamics on ecosystem-wide decision-making. Varying environmental conditions, we prove that the extracted GRNN computes input signals similar to natural decision-making process of the cell. Identifying of neural network behaviors in GRNs may lead to more accurate bacterial cell activity predictive models for many applications, including human health-related problems and agricultural applications. Further, this model can produce data on causal relationships throughout the network, enabling the possibility of designing tailor-made infection-controlling mechanisms. More interestingly, these GRNNs can perform computational tasks for bio-hybrid computing systems.
\end{abstract}


%
\IEEEpeerreviewmaketitle

\section{Introduction}

\IEEEPARstart{B}{acteria} are well-known for their capability to sense external stimuli, for complex information computations and for a wide range of responses \cite{Alm2006}. The microbes can sense numerous external signals, including a plethora of molecules, temperatures, pH levels, and the presence of other microorganisms \cite{Blair1995}. The sensed signals then go through the Gene Regulatory Network (GRN), where a large number of parallel and sequential molecular signals are collectively processed. The GRN is identified as the main computational component of the cell\cite{9858601}, which contains about 100 to more than 11000 genes (the largest genome identified so far belongs to \textit{Sorangium cellulosum} strain So0157-2) \cite{Land2015}. Despite the absence of neural components, the computational process through GRN allows the bacteria to actuate through various mechanisms, such as molecular production, motility, physiological state changes and even sophisticated social behaviors. Understanding the natural computing mechanism of cells can lead to progression of key areas of machine learning in bioinformatics, including prediction of biological processes, prevention of diseases and personalized treatment\cite{ravi2016deep}.  

\begin{figure}[t!]
    \centering
    \includegraphics[trim={0 30 0 0}, clip, width=0.9\linewidth]{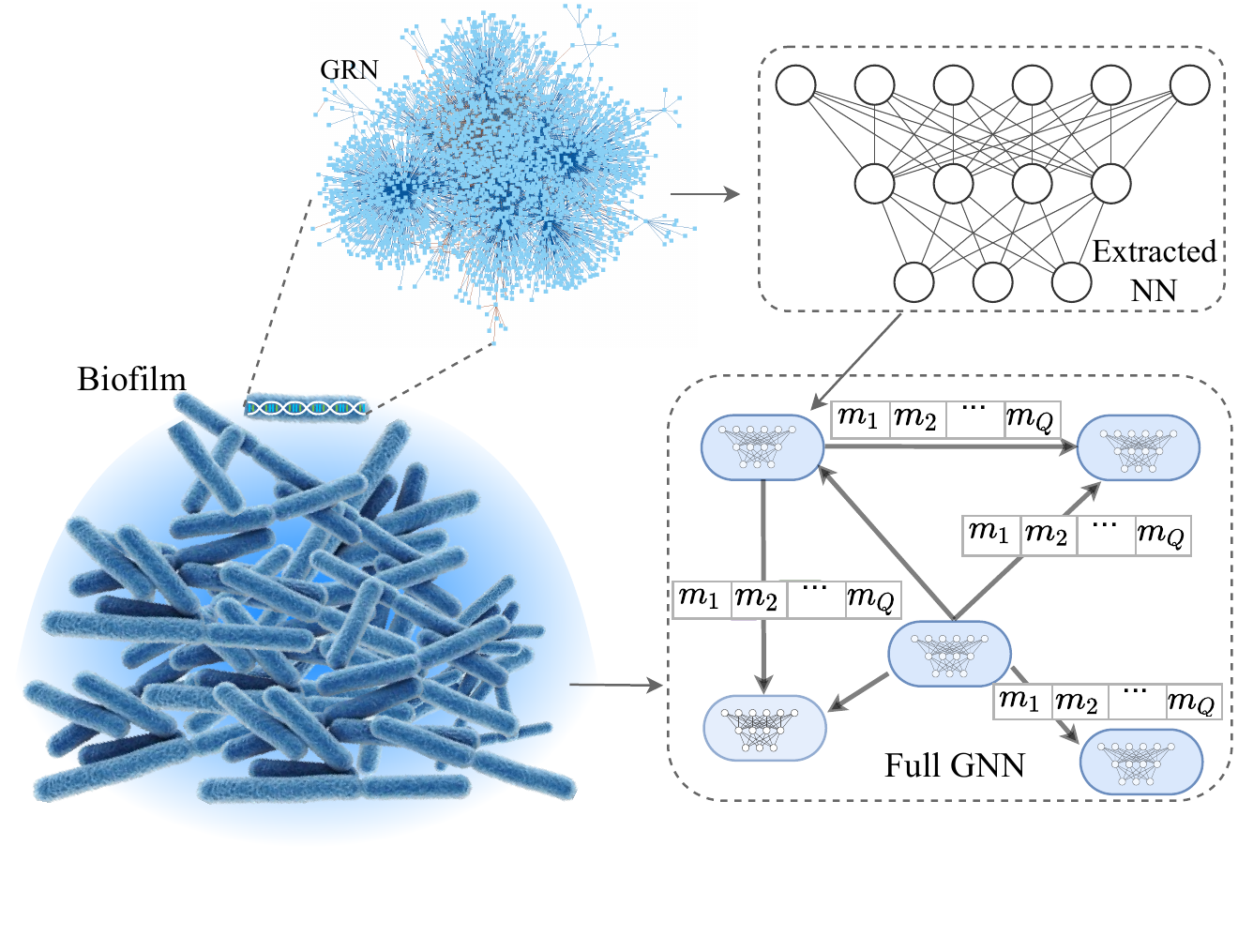}
    \caption{Illustration of the Gene Regulatory Neural Networks (GRNN) extraction and the implementation of the GNN to model the biofilm. The diffusion of molecules from one cell to another is modeled as a vector, where $m_q$ represents the concentration of the $q^{th}$ molecular signal. \vspace{-1.5em}}
    \label{fig:IntroDiagram}
\end{figure}

\begin{figure*}[t!]
\centering

\subfloat[\label{fig:TCSNet}]{
\includegraphics[trim={0 0 0 0},clip,width=0.3\textwidth]{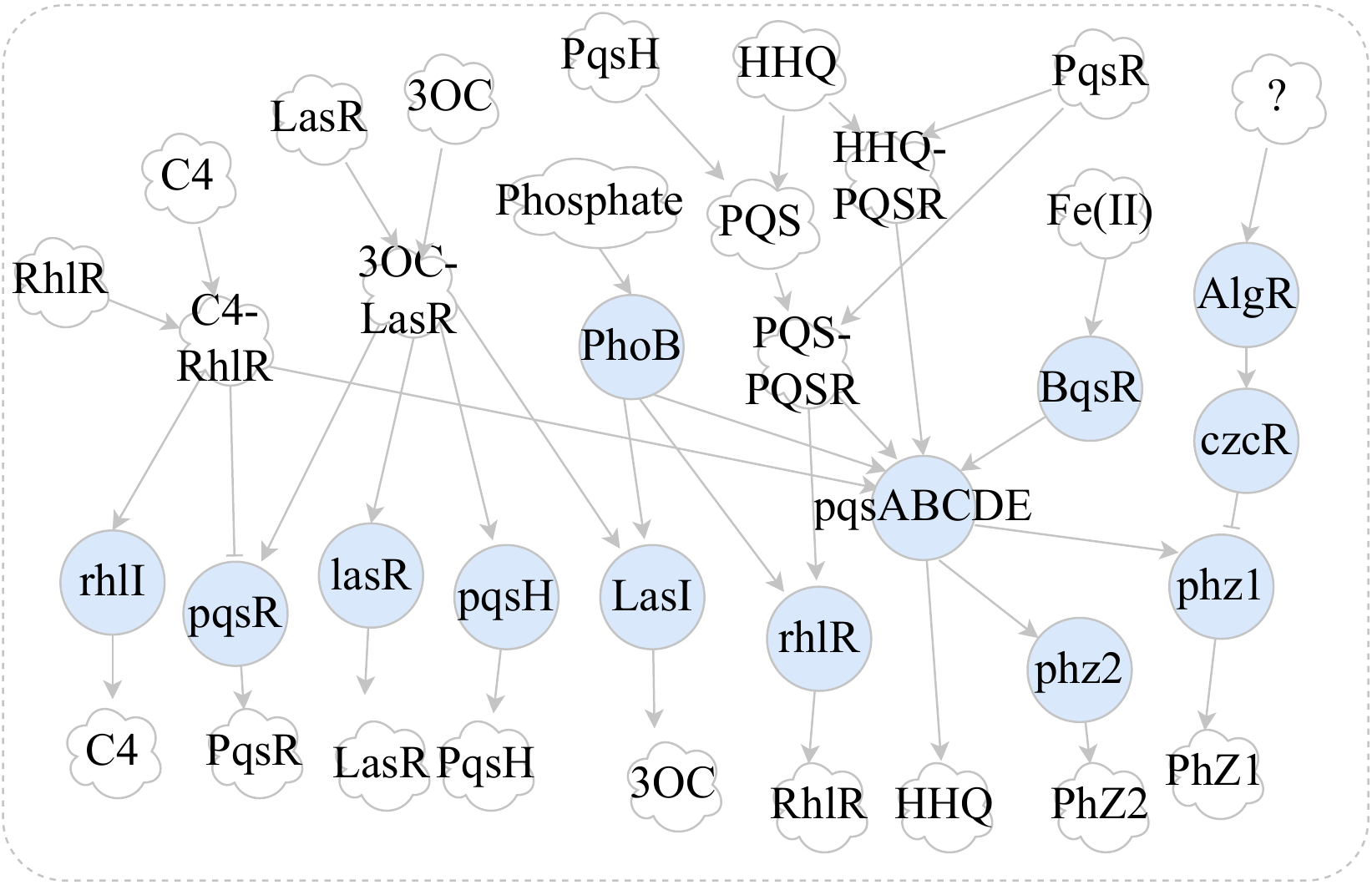}}
\subfloat[\label{fig:ANN}]{
\includegraphics[trim={0 0 0 0},clip,width=0.29\textwidth]{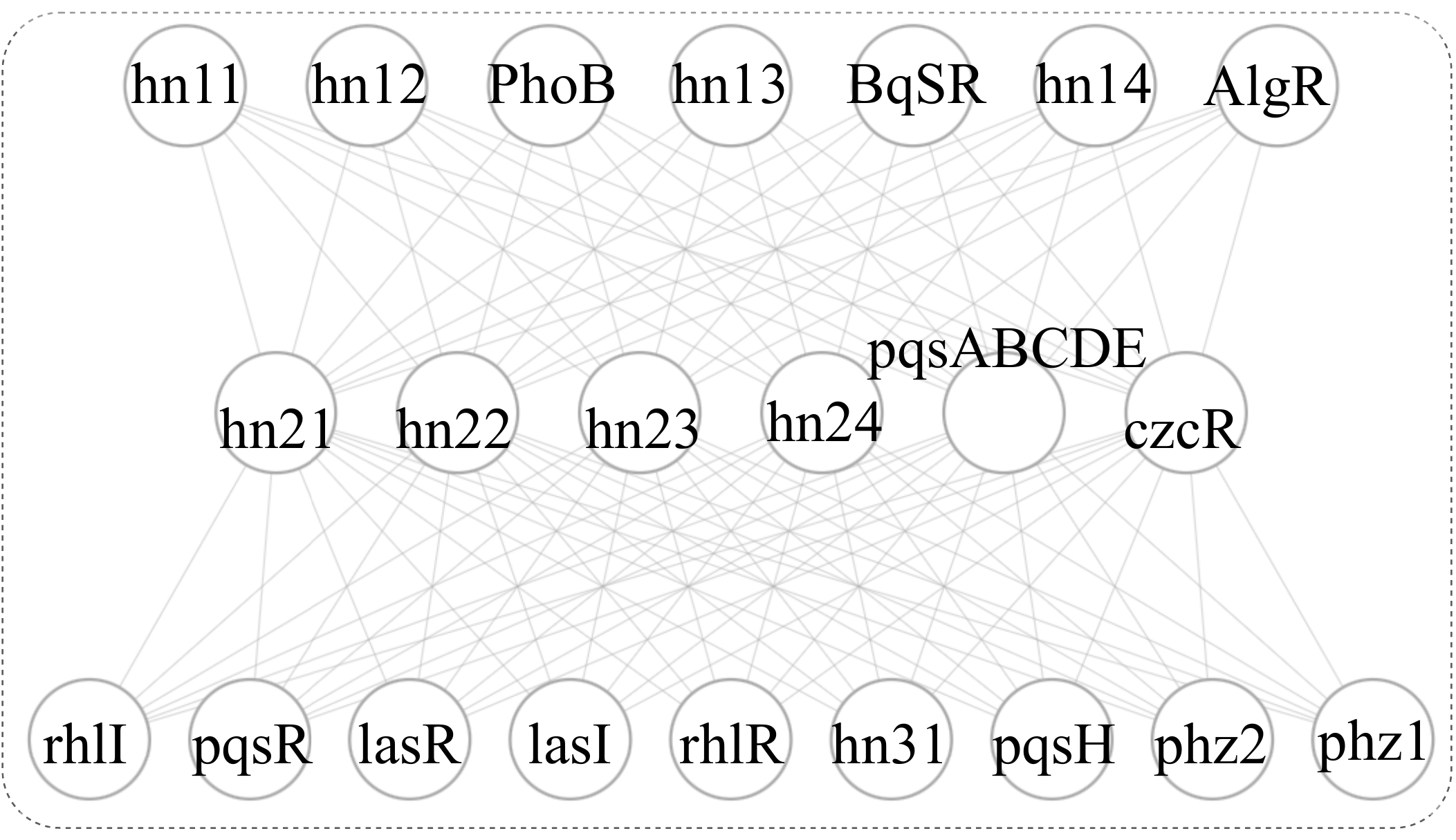}}
\subfloat[\label{fig:RealToModel}]{
\includegraphics[trim={0 0 0 0},clip,width=0.39\textwidth]{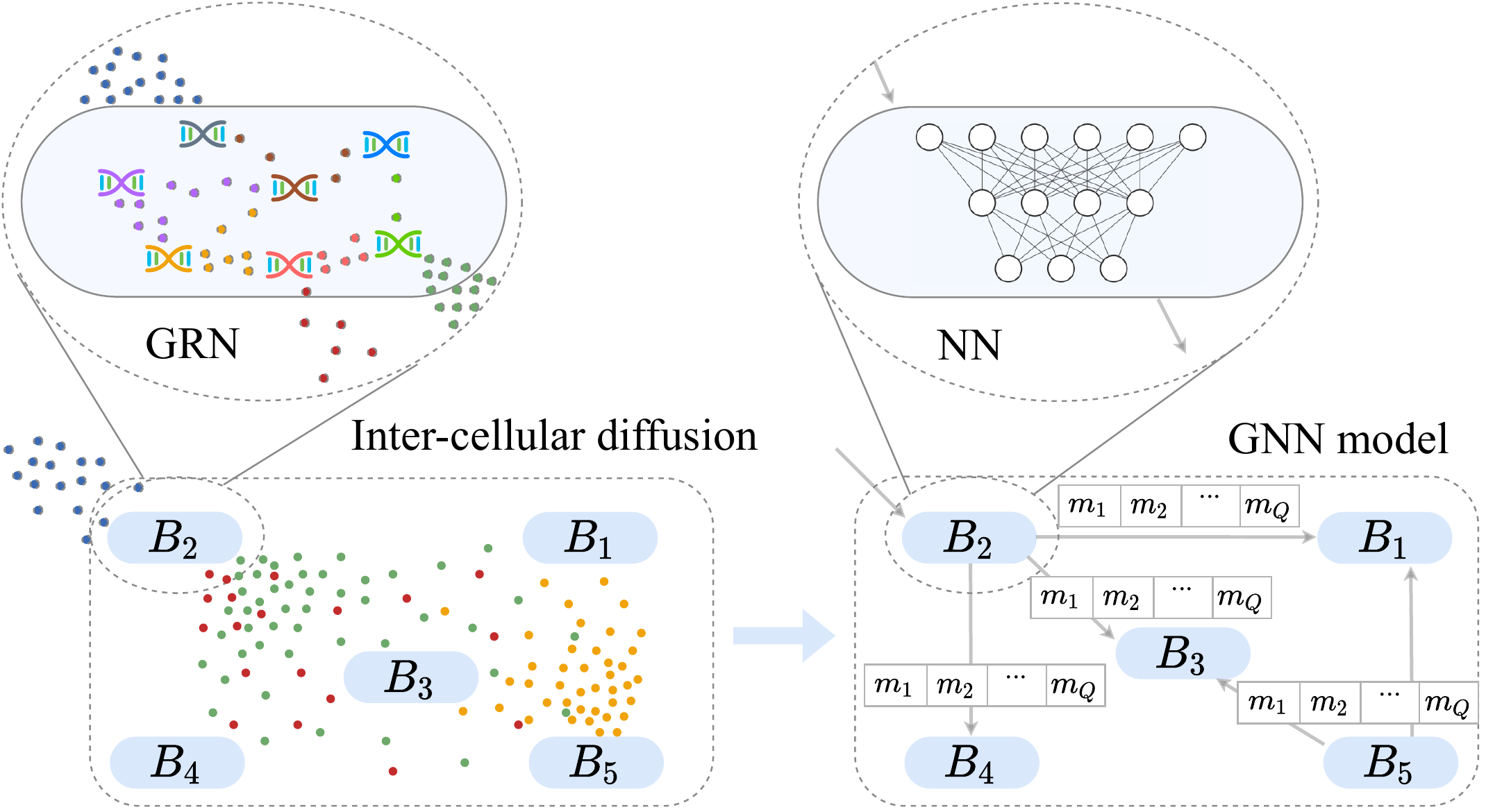}}
\caption{Extraction of a GRNN considering a specific sub-network of the GRN where a) is the two-component systems (TCSs) and QS network that is associated with the pyocyanin production, b) is the derived GRNN that is equipped with hypothetical nodes (\textbf{hn}s) without affecting its computation process to form a symmetric network structure and c) is the conversion of real biofilm to the suggested \textit{in-silico} model.  \vspace{-1em}}
\label{fig:GRNtoGNN}
\end{figure*}
 
Bacterial cells are equipped with various regulatory systems, such as single/two/multi-component systems including \emph{Quorum sensing} (QS), to respond to environmental stimuli. The receptors and transporters on cell membranes can react and transport extracellular molecules, which subsequently interact with respective genes. In turn, the GRN is triggered to go through a complex non-linear computational process in response to the input signals.
In the literature, it has been suggested that the computational process through the GRN of a bacterial cell comprises a hidden neural network (NN)-like architecture \cite{vohradsky2001neural, weaver1999modeling}. This indicates that, even though bacterial cells can be categorized as non-neural organisms, they perform neural decision-making processes through the GRN. This results in recent attention towards Molecular Machine Learning systems, where AI and ML are developed using molecular systems\cite{balasubramaniam2022realizing}. In these systems, several neural components can be identified in GRNs, in which genes may be regarded as computational units or neurons, transcription regulatory factors as weights/biases and proteins/second messenger Molecular Communications (MC) as neuron-to-neuron interactions. Owing to a large number of genes and the interactions in a GRN, it is possible to infer sub-networks with NN behaviors that we term Gene Regulatory Neural Networks (GRNN). The non-linear computing of genes results from various factors that expand through multi-omics layers, including proteomic, transcriptomic and metabolomic data (further explained in Section \ref{sec:cell_decision_making}). In contrast, the GRNN is a pure NN of genes with summarized non-linearity stemmed from multi-omics layers with weights/biases.

Identification of GRNNs can be used to model the decision-making process of the cell precisely, especially considering simultaneous multiple MC inputs or outputs. However, due to the limited understanding and data availability, it is still impossible to model the complete GRN with its NN-like behaviors. Therefore, this study uses a GRNN of \textit{Pseudomonas aeruginosa} that is associated with \textit{PhoR-PhoB} and \textit{BqsS-BqsR} two-component systems (TCSs) and three QS systems related to pyocyanin production as a use case to explore the NN-like behaviors. Although a single bacterium can do a massive amount of computing, they prefer living in biofilms. Hence, in order to understand the biofilm decision-making mechanism, we extend this single-cell computational model to an ecosystem level by designing an \textit{in-silico} single species biofilm with inter-cellular MC signaling as shown in Fig. \ref{fig:IntroDiagram}.

The contributions of this study are as follows:
\begin{itemize}
    \item \textbf{Extracting a GRNN:} Due to the complexity and insufficient understanding of the gene expression dynamics of the full GRN, we only focus on a sub-network associated with pyocyanin production (shown in Fig. \ref{fig:TCSNet}) to investigate the NN-like computational behavior of the GRN. Further, the genes of extracted sub-network are arranged following a NN structure that comprises input, hidden and output layers, as shown in Fig. \ref{fig:ANN}.
    
    \item \textbf{Modeling a biofilm as a GNN:} The GRNN only represents the single-cell activities. To model the biofilm-wide decision-making process, we use a Graph Neural Network (GNN). First, we create a graph network of the bacterial cell and convert it to a GNN by embedding each node with the extracted GRNN as the update function. Second, the diffusion-based MCs between bacterial cells in the biofilm are encoded as the message-passing protocol of the GNN, as shown in Fig. \ref{fig:RealToModel}.
    \item \textbf{Exploring the behaviors of the GRNN and intra-cellular MC dynamics to predict cell decisions:} The output of the GRNN is evaluated by comparing it with the transcriptomic and pyocyanin production data from the literature. Finally, an edge-level analysis of the GRNN is conducted to explore the causal relationships between gene expression and pyocyanin production.
\end{itemize}

This paper is organized as follows: Section \ref{sec:background} explains the background of bacterial decision-making in two levels: cellular-level in Section \ref{sec:cell_decision_making} and population-level in Section \ref{sec:biofilm_decision_making}, while the background on the \textit{P. aeruginosa} is introduced in Section \ref{sec:PA}. Section \ref{sec:System_design} is dedicated to explaining the model design of cellular and population levels. The results related to model validation and the intergenic intra-cellular signaling pattern analysis are presented in Section \ref{sec:Results} and the study is concluded in Section \ref{sec:Conclusion}.\vspace{-1em}

\section{Background}

As the model expands through single cellular and biofilm-wide decision-making layers, this section provides the background of how a bacterium uses the GRN to make decisions and how bacterial cells make decisions in biofilms. Moreover, we briefly discuss the cellular activities of the \textit{Pseudomonas aeruginosa} as it is the use case species of this study.\vspace{-1em}
\label{sec:background}

\subsection{Decision-Making Process of an Individual Cell}
\label{sec:cell_decision_making}
Prokaryotic cells are capable of sensing the environment through multiple mechanisms, including TCSs that have been widely studied and it is one of the focal points of this study. The concentrations of molecular-input signals from the extracellular environment influence the bacterial activities at the cellular and ecosystem levels \cite{unluturk2016impact}. Apart from the extracellular signals of nutrients, it is evident that the QS input signals have a diverse set of regulative mechanisms in biofilm-wide characteristics, including size and shape \cite{de2009quorum}. These input signals undergo a computational process through the GRN, exhibiting a complex decision-making mechanism. Past studies have explored and suggested this underpinning computational mechanism in a plethora of directions, such as using differential equations \cite{chen1999modeling} and probabilistic Boolean networks \cite{1046956} and logic circuit \cite{wang2015loregic}. All of these models mainly infer that the bacterial cells can make decisions not just based on the single input-output combinations, but they can integrate several incoming signals non-linearly to produce outputs.

\begin{figure}[t!]
    \centering
    \includegraphics[trim={5 0 0 0}, clip, width=0.9\linewidth]{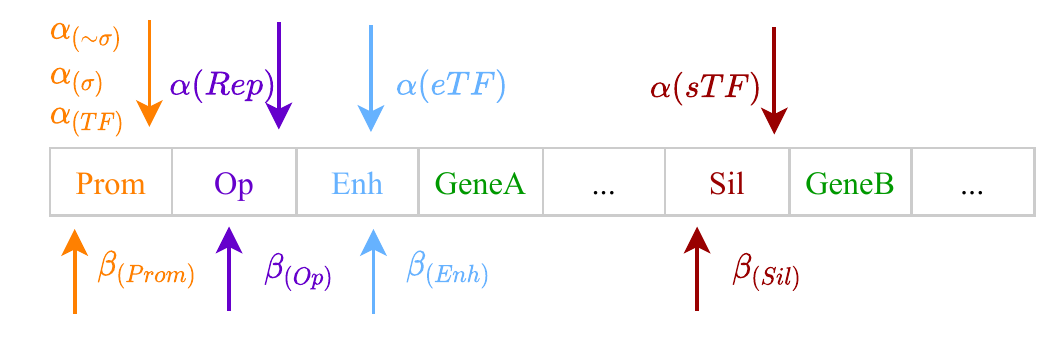}
    \caption{Illustration of gene expression regulators that are considered the weight influencers of the edges of GRNN. Here, the $\alpha_{(\sigma)}$, $\alpha_{(\sim \sigma)}$, $\alpha_{(TF)}$, $\alpha_{(Rep)}$, $\alpha_{(eTF)}$ and $\alpha_{(sTF)}$ are relative concentrations of sigma factors, anti-sigma factors, transcription factors (TFs), repressors, enhancer-binding TFs and silencer-binding TFs respectively. Moreover, $\beta_{(Prom)}$, $\beta_{(Op)}$, $\beta_{(Enh)}$, and $\beta_{(Sil)}$ are the binding affinities of the promoter, operator, enhancer and silencers regions respectively.\vspace{-1em}}
    \label{fig:WeightInfluencers}
\end{figure}

\begin{figure}[t!]
\centering

\subfloat[\label{fig:37CNN}]{
\includegraphics[trim={0 0 0 0},clip,width=0.8\linewidth]{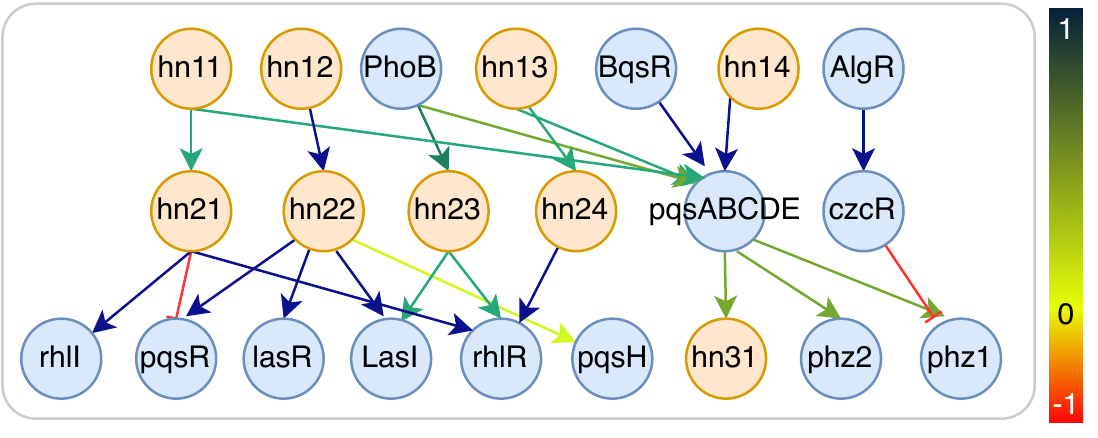}}

\subfloat[\label{fig:30CNN}]{
\includegraphics[trim={0 0 0 -10},clip,width=0.8\linewidth]{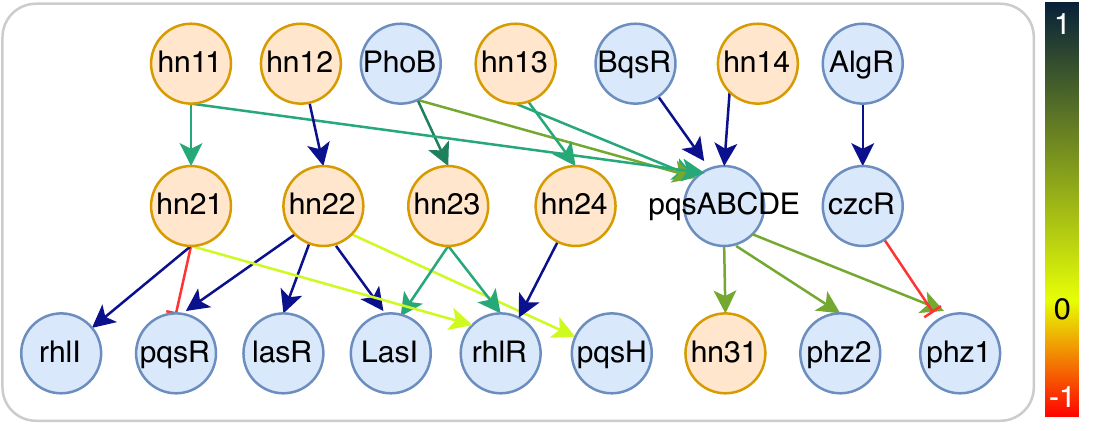}}
\caption{Two GRNN setups with different weights associated with two environmental conditions. a) is the relative weight setup of \textit{P. aeruginosa} cell in 37$\,^{\circ}$C and b) is in 30$\,^{\circ}$C. \vspace{-1.2em}}
\label{fig:DiffWeightsInDiffWeights}
\end{figure}

The studies that focus on differences in gene expression levels suggest that a hidden weight behavior controls the impact of one gene on another \cite{vohradsky2001neural}. This weight behavior emerges through several elements, such as the number of transcription factors that induce the expression, the affinity of the transcription factor binding site, and machinery such as thermoregulators and enhancers/silencers \cite{grosso2014regulation, ishihama2012prokaryotic}. Fig. \ref{fig:WeightInfluencers} depicts a set of factors influencing the weight between genes. The weight of an edge between two genes has a higher dynamicity as it is combinedly determined by several of these factors.
Based on environmental conditions, the GRN of the bacterial cell adapts various weights to increase the survivability and repress unnecessary cellular functions to preserve energy. An example of such regulation is shown in Fig. \ref{fig:DiffWeightsInDiffWeights} where a \textit{P. aeruginosa} cell uses a thermoregulator to regulate the QS behaviors. Fig. \ref{fig:37CNN} has a set of relative weights based on cellar activities in an environment at 37$\,^{\circ}$C, while Fig. \ref{fig:30CNN} represents weights at 30$\,^{\circ}$C. The weights between the \textbf{hn21} and \textit{rhlR} are different in two conditions, and these cellar activities are further explained in \cite{grosso2014regulation}.\vspace{-1.0em}
 
\subsection{Biofilm Decision-Making}
\label{sec:biofilm_decision_making}

Even though an individual cell is capable of sensing, computing, and actuating, the majority of bacterial cells live in biofilms, where the survivability is significantly increased compared to their planktonic state. 
Biofilm formation can cause biofouling and corrosion in water supply and industrial systems \cite{fletcher1994bacterial}. However, biofilms formation can be desirable in many situations, for example, bioreactors in wastewater treatment \cite{qureshi2005biofilm}, bioremediation of contaminated groundwater \cite{kao2001development, 8468188}, where biofilms serve as platforms for biogeochemical reactions.
A massive number of factors can influence biofilm formation, including substratum surface geometrical characteristics, diversity of species constituting the biofilm, hydrodynamic conditions, nutrient availability, and especially communication patterns \cite{Goller2008} where the TCS and QS play significant roles. A TCS comprises a \emph{histidine kinase} that is the sensor for specific stimulus and a cognate response regulator that initiates expressions of a set of genes \cite{mascher2006stimulus}. Hence, in each stage, essential functions can be traced back to their gene expression upon a response to the input signals detected by bacterial cells. For instance, in the first stage of biofilm formation, the attachment of bacteria to a surface is associated with sensing a suitable surface and altering the activities of the flagella. In the next stage, \emph{rhamnolipids} production is associated with ferric iron $\text{Fe}^{3+}$ availability in the environment, mostly sensed through \emph{BqsS-BqsR} TCSs. Further, $\text{Fe}^{3+}$ was identified as a regulator of \emph{pqsA}, \emph{pqsR}, and \emph{pqsE} gene expressions that are associated with the production of two critical components for the formation of microcolonies: eDNA and EPS\cite{passos2017update}. Similarly, in the final stage, the dispersion process can also be traced back to a specific set of gene regulations, including \textit{bdlA} an \textit{rbdA} \cite{toyofuku2016environmental, Rumbaugh2020}. An understanding of the underlying decision-making process of bacteria may enable us to control their cellular activities.\vspace{-1.0em} 
\subsection{Pseudomonas Aeruginosa}
\label{sec:PA}
The main reason for selecting \textit{P. aeruginosa} in this work lies in its alarming role in human health. For example, this species is the main cause of death in cystic fibrosis patients \cite{driscoll2007epidemiology}. \textit{P. aeruginosa} is a gram-negative opportunistic pathogen with a range of virulence factors, including pyocyanin and cytotoxin secretion\cite{Strateva2011}. These secreted molecules can lead to complications such as respiratory tract ciliary dysfunction and induce proinflammatory and oxidative effects damaging the host cells \cite{lau2005modulation}. The biofilms are being formed on more than 90\% endotracheal tubes implanted in patients who are getting assisted ventilation, causing upper respiratory tract infections \cite{Pericolini2018}. In addition, another important reason for targeting \textit{P. aeruginosa} is the data availability for the GRN structure \cite{GalnVsquez2020}, pathways \cite{Caspi2019}, genome \cite{Winsor2015}, transcriptome \cite{Barrett2012} and data from mutagenesis studies \cite{Kulasekara2004, Kilmury2018}. Compared to the complexity of the GRN, the amount of data and information available on the gene-to-gene interactions and expression patterns is insufficient to develop an accurate full \emph{in-silico} model. Therefore, we chose a set of specific genes that are associated with QS, TCS, and pyocyanin production.\vspace{-1em}

\section{System design}
\label{sec:System_design}
This section explains the system design in two main phases, extracting a NN-like architecture from the GRN targeting the set of genes and creating a model of the biofilm ecosystem.\vspace{-1em}

\subsection{Extracting Natural Neural Network from GRN}


We first fetch the structure of the GRN graph from RegulomePA \cite{GalnVsquez2020} database that contains only the existence of interactions and their types (positive or negative). As the next step, using information from the past studies \cite{venturi2006regulation, francis2017two, sultan2021roles, chambonnier2016hybrid}, we identified the genes involved in the \textbf{Las}, \textbf{Rhl} and \textbf{PQS} QS systems, \textit{PhoR-PhoB} and \textit{BqsS-BqsR} TCSs, and pyocyanin production to derive the sub-network of GRN as shown in Fig \ref{fig:TCSNet}. 
We further explored the expression dynamics using transcriptomic data  \cite{kreamer2012bqsr, huang2019integrated} where we observed the non-linearity in computations that are difficult to capture with existing approaches such as logic circuits, etc. \cite{vohradsky2001neural}, making the NN approach more suitable. However, a NN model with a black box that is trained on a large amount of transcriptomic data records to do computations similar to the GRN has a number of limitations, especially in understanding the core of the computational process\cite{akay2019deep}. Our model does not use a conventional NN model; instead, we extract a NN from the interaction patterns of the GRN, which we consider a pre-trained GRNN. In this sub-network, we observed that the lengths of expression pathways are not equal. For example, the path from \textit{PhoR-PhoB} to the \textit{phz2} gene has two hops, but the path from the \textit{BqsS-BqsR} system to the \textit{rhlR} gene only has one hop. The extracted network has the structure of a random NN. Hence, we transform this GRNN to Gene Regulatory Feedforward Neural Network by introducing hypothetical nodes (\textbf{hn}s) that do not affect the behaviors of the GRNN as shown in Fig \ref{fig:ANN}. In this transformation, we decide the number of hidden layers based on the maximum number of hops in gene expression pathways. In our network, the maximum number of hops is two, which determines the number of hidden layers as one, and then the number of hops of all the pathways is leveled by introducing \textbf{hn}s. If a \textbf{hn} is introduced between a source and target genes, the edge weights from the source node to the \textbf{hn} and from \textbf{hn} to the target node are made ``1'' so that the \textbf{hn} does not have an influence on the regulation of genes. Moreover, if a gene does not induce another in the network, the weight of the edge between that pair is made ``0''.

\begin{figure}[t!]
\centering
\includegraphics[width=0.8\linewidth]{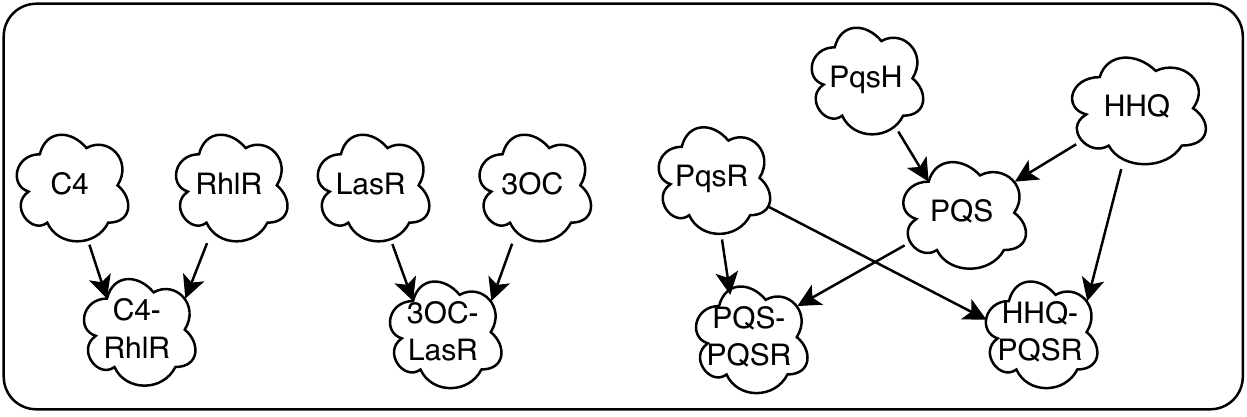}
\caption{Illustrations of intra-cellular metabolite interaction.}
\label{fig:Intracellular}
\end{figure}

Here, we summarize multiple factors of interaction into a weight that determines the transcriptional regulation of a particular gene. This regulation process occurs when the gene products get bound to the promoter region of another, influencing the transcriptional machinery. Hence, we observe this regulation process of a target gene as a multi-layered model that relies on the products of a set of source genes, the interaction between gene products, and the diffusion dynamics within the cell. Creating a framework to infer an absolute weight value using all the above factors is a highly complex task. In order to infer weight, one method is to train a NN model with the same structure as the GRN using a series of transcriptomic data. However, this approach also has numerous challenges, such as the lack of a sufficient amount of data in similar environments.

Therefore, we estimate a set of relative weights based on genomic, transcriptomic, and proteomic explanations of each interaction from the literature. The weights were further fine-tuned using the transcriptomics data. A relative weight value of an edge can be considered a summarizing of multi-layer transcriptional-translation to represent the impact of the source gene on a target gene.

In this computational process, we identify another layer of interactions that occur within the cell. The produced molecules by the considered TCs network go through a set of metabolic interactions that are crucial for the functionality of the cell. Since our primary goal is to explore the NN behaviors of GRN, we model these inter-cellular chemical reactions as a separate process, keeping the gene-to-gene interactions and metabolic interactions in two different layers. To model the complete pyocyanin production functionality of the cell, we use the inter-cellular molecular interactions shown in Fig \ref{fig:Intracellular}. Here, RhlR is a transcriptional regulator of \textit{P. aeruginosa} that forms a complex by getting attached to its cognate inducer C4-HSL and then binds to the promoter regions of relevant genes \cite{lamb2003functionaldomains}. Similarly, LasR transcriptional regulator protein and 3-oxo-C12-HSL (3OC), and PqsR with PQS and HHQ form complexes and get involved in the regulation of a range of genes \cite{pearson1997roles, wade2005regulation}. Further, C10H10O6 in the environment are converted by the \textit{P. aeruginosa} cells in multiple steps using the products of the GRNN we consider. First, C10H10O6 is converted into phenazine-1-carboxylic using the enzymes of \textit{pqsABCDEFG} genes. Later,  phenazine-1-carboxylic was converted into 5-Methylphenazine-1-carboxylate, and finally, 5-Methylphenazine-1-carboxylate into Pyocyanin by PhzM and PhzS, respectively \cite{nadal2012multiple}.

Molecular accumulation within a bacterial cell can be considered its memory module where certain intra-cellular interactions occurs. Therefore, we define an internal memory matrix $IM$ as,
\begin{equation} 
\begin{split}
&\mathbf{IM}^{(t)} =
\begin{blockarray}{ccccc}
 & im_{1} & im_{2} & ... & im_{J}\\
\begin{block}{c(cccc)}
 B_{1} & C_{(1,im_1)}^{(t)}  & C_{(1,im_2)}^{(t)}  &... & C_{(1,im_J)}^{(t)}\\ 
 B_{2} & C_{(2,im_1)}^{(t)}  & C_{(2,im_2)}^{(t)}  &... & C_{(2,im_J)}^{(t)} \\ 
 \vdots  &\vdots  &\vdots    &\ddots  &\vdots \\ 
 B_{P} & C_{(P,im_1)}^{(t)}  & C_{(P,im_2)}^{(t)}  &... & C_{(P,im_J)}^{(t)}\\
\end{block}
\end{blockarray}
,\end{split}
\end{equation}

where the concentration of the internal molecule $im_j$ is $C^{(t)}_{(i,im_j)}$.

GRNN process molecular signals from the environment and other cells. Hence, we used the approach of GNN as a scalable mechanism to model the MCs and biofilm wide decision-making process. The extreme computational power demand of modeling the diffusion-based MCs of each cell is also avoided by using this approach. \vspace{-1em}

\subsection{Graph Neural Network Modeling of Biofilm}
\begin{figure}[t!]
\centering

\subfloat[\label{fig:GraphNetwork}]{
\includegraphics[width=0.6\linewidth]{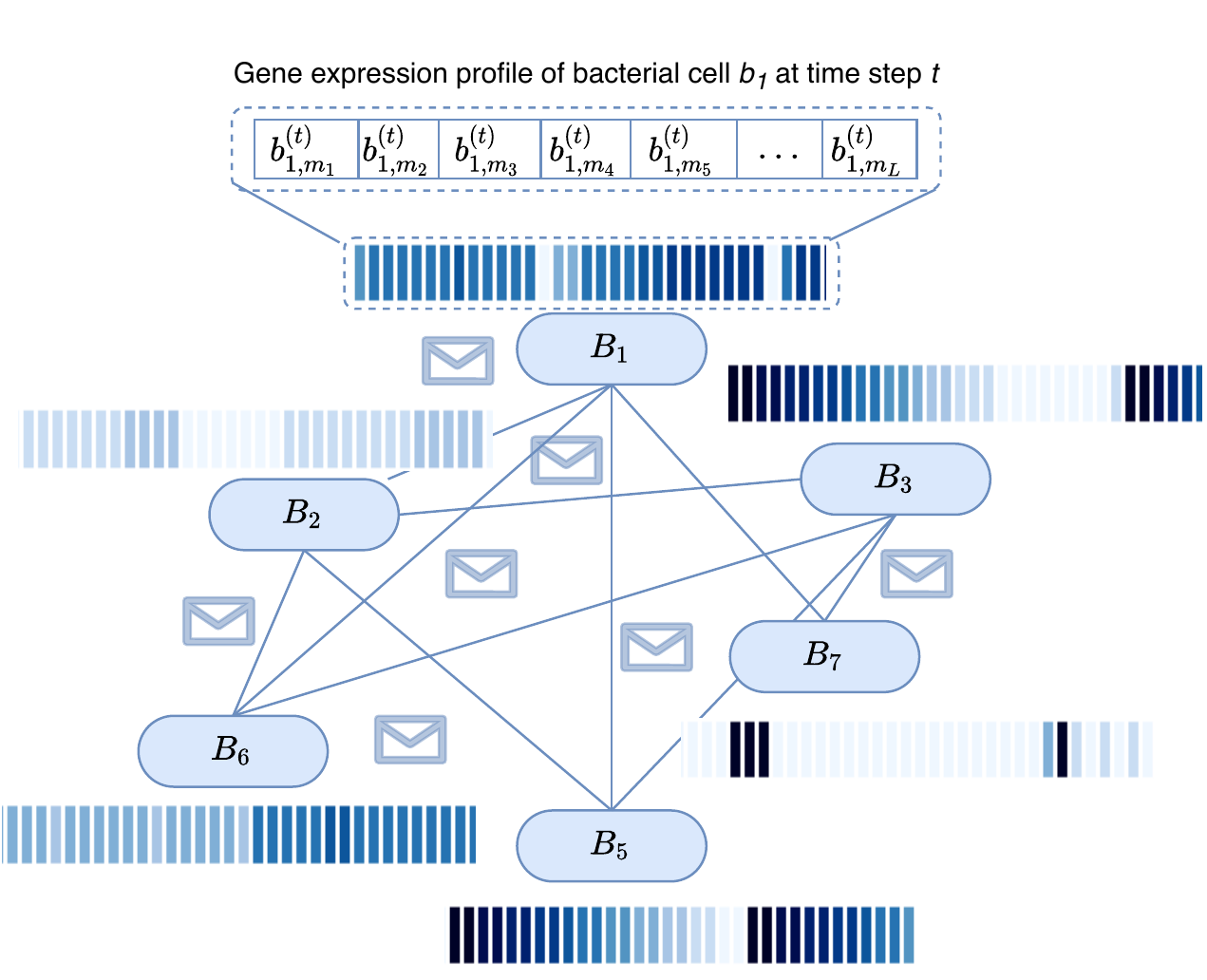}}

\subfloat[\label{fig:LearningProcess}]{
\includegraphics[width=0.9\linewidth]{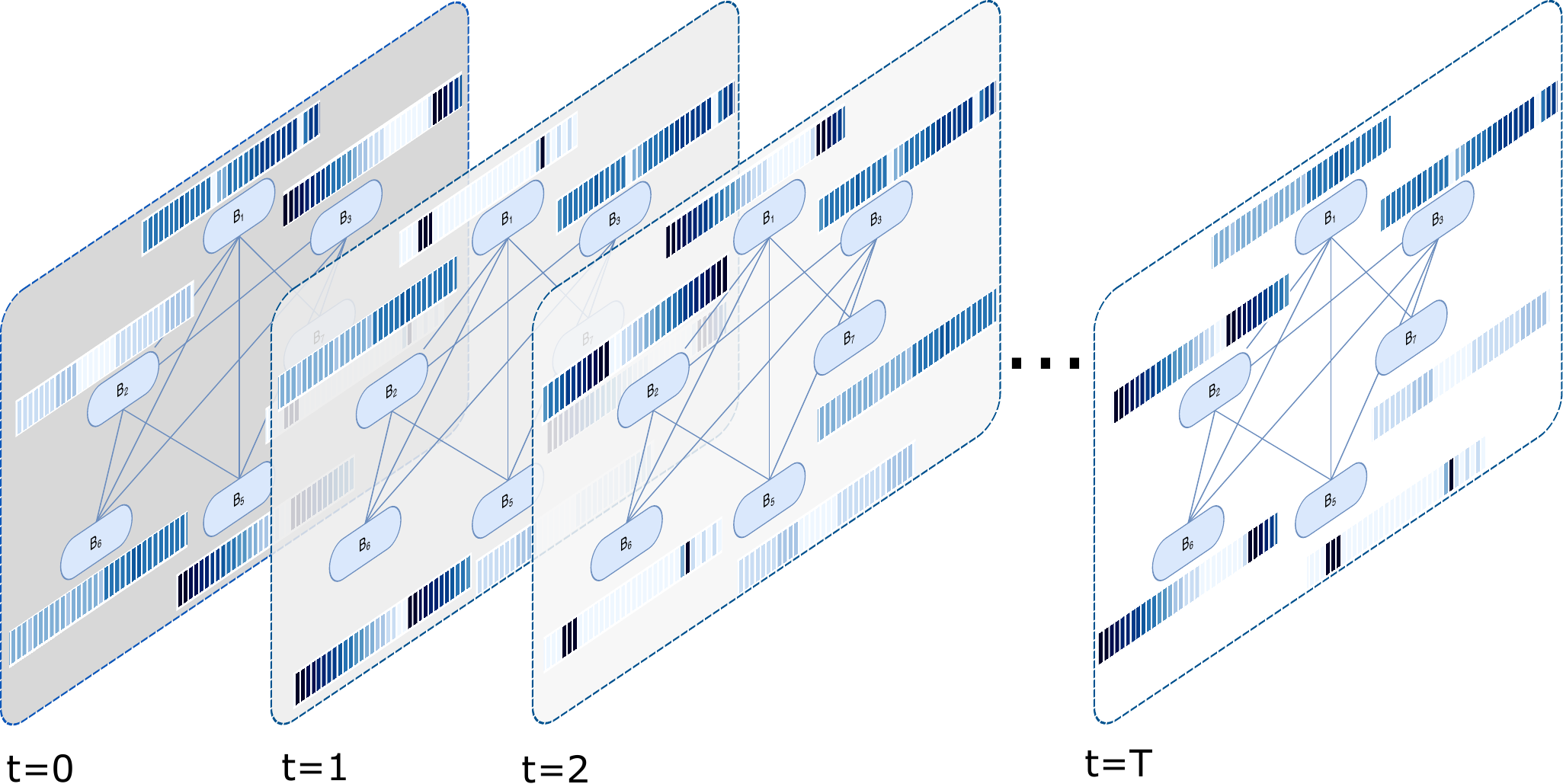}}
\caption{Illustration of the GNN components where a) is a snapshot of the bacterial network that has the gene expression profile as the feature vector. Further, this gene expression pattern of a cell is encoded to a message of secreted molecules where MC plays a crucial role. Moreover, b) shows the temporal behavior of the GNN, that the output of one graph snapshot influences the next.\vspace{-1.5em}}
\label{fig:FullGNN}
\end{figure}

First, the biofilm is created as a graph network of bacterial cells where each node is a representation of a cell, and an edge between two nodes is a MC channel. We convert the graph network into a Graph Neural Network (GNN) in three steps: 1) embedding the extracted GRNN of pyocyanin production into each node as the update function, 2) encoding the diffusion-based cell-to-cell MC channels as the message passing scheme, and 3) creating an aggregation function at the reception of molecular messages by a node as shown in Fig. \ref{fig:FullGNN}. Next, we define feature vectors of each node of the GNN to represent the gene expression profile of the individual cell at a given time. Subsequently, considering $L$ is the number of genes in the GRNN, $P$ is the number of bacterial cells in the biofilm and $b_{(i,g_l)}^{(t)}$ is the expression of gene $g_l$ by the bacteria $B_i$, we derive the following matrix $\mathbf{FV}^{(t)}$ that represents all the feature vectors of the GNN at time $t$.\vspace{-1em}

\begin{equation} 
\mathbf{FV}^{(t)} = 
\begin{blockarray}{ccccc}
 & g_{1} & g_{2} & ... & g_{L}\\
\begin{block}{c(cccc)}
 B_{1} & b_{(1,g_1)}^{(t)}  & b_{(1,g_2)}^{(t)}  &... & b_{(1,g_L)}^{(t)}\\ 
 B_{2} & b_{(2,g_1)}^{(t)}  & b_{(2,g_2)}^{(t)}  &... & b_{(2,g_L)}^{(t)} \\ 
 \vdots  &\vdots  &\vdots    &\ddots  &\vdots \\ 
 B_{P} & b_{(P,g_1)}^{(t)}  & b_{(P,g_2)}^{(t)}  &... & b_{(P,g_L)}^{(t)}\\
\end{block}
\end{blockarray}
\end{equation} 
 
\begin{figure}[t!]
    \centering
    \includegraphics[trim={0 5 0 15}, clip, width=\linewidth]{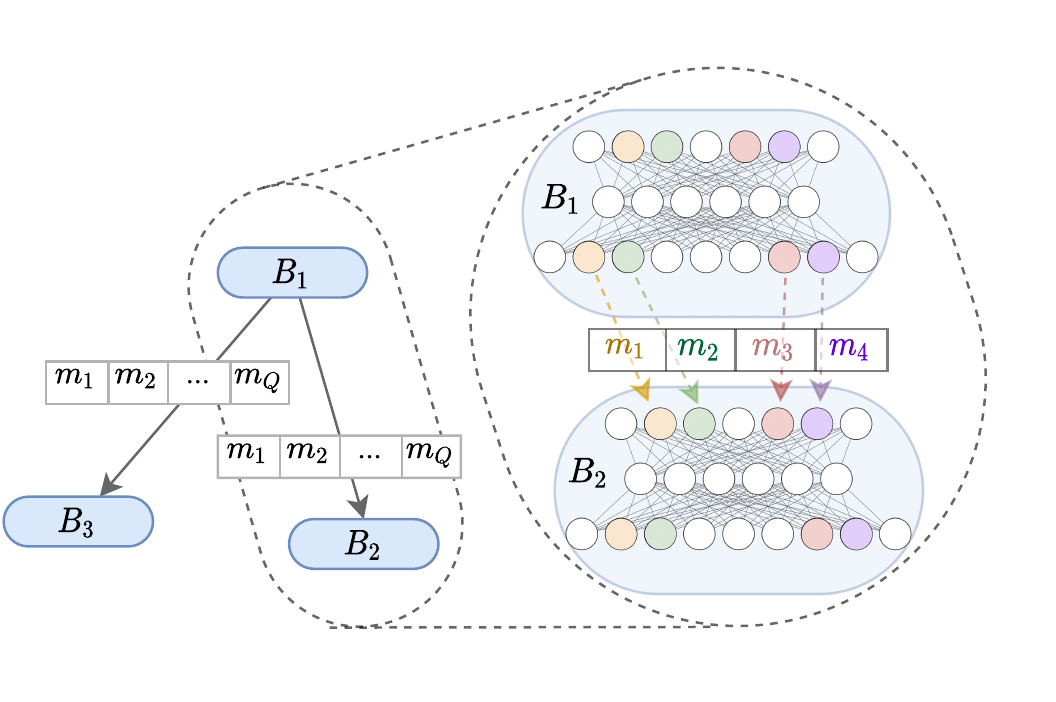}
    \caption{The process of one GRNN outputs reaching another GRNN as molecular messages.\vspace{-1.3em}}
    \label{fig:NNtoNN}
\end{figure}

The computational output of the GRNN of each node results in the secretion of a set of molecules that are considered messages in our GNN model as illustrated in the Fig. \ref{fig:NNtoNN}.

When the number of molecular species considered in the network is $Q$ and output $m_q$ molecular message from bacterial cell $B_i$ at TS $t$ is $msg_{(i,m_q)}^{(t)}$, we derive the matrix 
\begin{equation} 
\begin{split}
&\mathbf{MSG}^{(t)} = \\
&
\begin{blockarray}{ccccc}
 & m_{1} & m_{2} & ... & m_{Q}\\
\begin{block}{c(cccc)}
 B_{1} & msg_{(1,m_1)}^{(t)}  & msg_{(1,m_2)}^{(t)}  &... & msg_{(1,m_Q)}^{(t)}\\ 
 B_{2} & msg_{(2,m_1)}^{(t)}  & msg_{(2,m_2)}^{(t)}  &... & msg_{(2,m_Q)}^{(t)} \\ 
 \vdots  &\vdots  &\vdots    &\ddots  &\vdots \\ 
 B_{P} & msg_{(P,m_1)}^{(t)}  & msg_{(P,m_2)}^{(t)}  &... & msg_{(P,m_Q)}^{(t)}\\
\end{block}
\end{blockarray}
.\end{split}
\end{equation}
\begin{table*}[t!]
\caption{Parameters utilised in the system development}
\label{table:Parameters}
\begin{threeparttable}
\begin{tabular}{lrl}
\hline
\textbf{Parameter}                                       & \multicolumn{1}{r}{\textbf{Value}}                         & \textbf{Description}  \\                   \hline
No. of cells                    & 2000                      & The number of cells is limited due to the memory availability of the server.         \\
No. of genes                    & 13                        & \vtop{\hbox{\strut The network only consists of the gene that are directly associated with QS, \textit{PhoR-PhoB} and \textit{BqsS-BqsR}}\hbox{\strut TCSs, and pyocyanin production.}}         \\
No. internal memory molecules   & 16                        &  The set of molecules that involved in QS, \textit{PhoR-PhoB} and \textit{BqsS-BqsR} TCSs,and pyocyanin production.         \\
No. messenger molecules         & 4                         & The number of molecules that were exchanged between cells in the sub network.         \\
Dimensions of the environment   & 20x20x20$\mu m$    & \vtop{\hbox{\strut The dimensions were fixed considering the average sizes of \textit{P. aeruginosa} biofilms and computational }\hbox{\strut demand of the model.}}         \\
Duration                        & 150 TSs            & \vtop{\hbox{\strut The number of TSs can be modified to explore the cellular and ecosystem level activities. For this }\hbox{\strut experiment we fixed a TS to represent 30mins.}}         \\
No. iterations per setup        & 10                        & \vtop{\hbox{\strut Considering the stochasticity ranging from the  gene expression to ecosystem-wide communications, the }\hbox{\strut experiments were iterated 10 times.}}         \\
\hline
\end{tabular}
\end{threeparttable}
\end{table*}
Further, we use a static diffusion coefficients vector 
\begin{equation}
    \mathbf{D} = \{ D_{m_1},D_{m_2},...,D_{m_Q}\},
\end{equation}
where $D_{m_q}$ is diffusion coefficient of molecular species $m_q$.

We define another matrix $\mathbf{ED}$ that contains the euclidean distances between bacterial cells in the biofilm as follows
\begin{equation} 
\mathbf{ED} = 
\begin{blockarray}{ccccc}
 & B_{1} & B_{2} & ... & B_{P}\\
\begin{block}{c(cccc)}
 B_{1} & d_{(1,1)}  & d_{(1,2)}  &... & d_{(1,P)}\\ 
 B_{2} & d_{(2,1)}  & d_{(2,2)}  &... & d_{(2,P)} \\ 
 \vdots  &\vdots  &\vdots    &\ddots  &\vdots \\ 
 B_{P} & d_{(P,1)}  & d_{(P,2)}  &... & d_{(P,P)}\\
\end{block}
\end{blockarray}
\end{equation}
where $d_{i,j}$ is the euclidean distance between the $i^{th}$ and $j^{th}$ cells.

The feature vector of $i^{th}$ bacterial cell at the TS $t+1$ is then modeled as,

\begin{equation}
\label{eq:GNN_main_model}
    \mathbf{FV}_{i}^{(t+1)}=GRNN_i(\mathbf{MSG}_{i}^{(t)}+S^{(t)}_{i})
\end{equation}
where $MSG_i^{(t)}$ is the message generated by the same cell in the previous TS. The $GRNN_i$ is the extracted GRNN that is the update function in the GNN learning process and $\mathbf{S}^{(t)}_{i} =  \mathbf{R}_i^{(t)}+\mathbf{K}_i^{(t)}$,
is the aggregate function. In the aggregation component, the $R_i^{(t)}$ is the incoming signals from peer bacterial cells and $K_{(i:m_q)}^{(t+1)}$ is the external molecule input vector at the location of $B_i$ and the TS $t$ that is expressed as
\begin{equation}
    \mathbf{K}_i^{(t+1)} = \big \{ K_{i:m_1}^{(t+1)},K_{i:m_2}^{(t+1)},...,K_{i:m_Q}^{(t+1)} \big \}.
\end{equation}

In order to compute $\mathbf{R}_i^{(t+1)}$, we use a matrix $\mathbf{Y}_i$; $\mathbf{Y}_i = \stackrel{\leftrightarrow}{1}_{[Q\times 1]} \times \mathbf{ED}_i$,
where $\stackrel{\leftrightarrow}{1}_{[Q\times 1]}$ is an all-ones matrix of dimension $Q \times 1$. The $\hat{g}$ matrix is then defined as follows,
\begin{equation} 
\begin{split}
&\mathbf{\hat{g}}(\mathbf{D}^ \intercal, \mathbf{Y}, t) = \\
&\begin{bmatrix}
g(D_{m_1},d_{(i,1)},t)&  g(D_{m_1},d_{(i,2)},t)&  ...& g(D_{m_1},d_{(i,P)},t) \\ 
g(D_{m_2},d_{(i,1)},t)&  g(D_{m_2},d_{(i,2)},t)&  ...& g(D_{m_2},d_{(i,P)},t)\\ 
\vdots &  \vdots &  \ddots & \vdots\\ 
g(D_{m_Q},d_{(i,1)},t)&  g(D_{m_Q},d_{(i,2)},t)&  ...& g(D_{m_Q},d_{(i,P)},t)
\end{bmatrix}.
\end{split}
\end{equation}

In the above matrix, $g(D_{m_l},d_{(i,j)},t)$ is the Green's function of the diffusion equation as shown below,
\begin{equation}
    \hat{G}(D_{m_l},d_{(i,j)},t) = \frac{1}{\left ( 4\pi D_{m_l}
t \right )^{\frac{3}{2}}}\exp\left ( -\frac{d_{(i,j)}^2}{4D_{m_l}t}  \right ).
\end{equation}

Further, the incoming signal vector $\mathbf{R}_i^{(t+1)}$ is denoted as below,
\begin{equation}
    \mathbf{R}_i^{(t+1)} = diag \big ( \mathbf{\hat{g}}(\mathbf{D}^ \intercal, \mathbf{Y}, t) \times \mathbf{MSG}^{(t)} \big ).
\end{equation}

Further, we equip our model with a 3-D environment to compensate for the noise element and external molecule inputs. Environment-layer is designed as a 3-D grid of voxels that can store precise information on external nutrients (similarly to our previous model in \cite{9705067}). The diffusion of nutrient molecules through the medium is modeled as a random-walk process. 
This layer allows us to enrich the model with the dynamics of nutrient accessibility of bacterial cells due to diffusion variations between the medium and the Extracellular Polymeric Substance (EPS). 

The bacterial cells in the ecosystem also perform their own computing tasks individually, resulting in a massively parallel processing framework. Hence, we use the python-cuda platform to make our model closer to the parallel processing architecture of the biofilm, where we dedicate a GPU block for each bacterial cell and the threads of each block for the matrix multiplication of the GRNN computation associated with the particular cell. Additionally, due to the massive number of iterative components in the model, the computational power demand faces significant challenges with serial programming making parallelization the best match for the model.\vspace{-1em}
\section{Simulations}
\label{sec:Results}

In this section, we first explain the simulation setup and then discuss the results of gene expression and molecular production dynamics to prove the accuracy of the extracted GRNN, emphasizing that it works similarly to the real GRN. Later, we use computing through the GRNN to explain certain activities of the biofilm.\vspace{-1em}

\subsection{Simulation Setup}

As our interest is to investigate the NN-like computational process, we do not model the formation process of the biofilm, but we only remodel a completely formed biofilm and disregarding the maturation and dispersion stages. In this model, we consider the biofilm as a static 3-D structure of bacterial cells. Hence, we first place bacterial cells randomly in the model in a paraboloid shape using the equation, $ z<\frac{x^2}{5} + \frac{y^2}{5} +20$ where $x$, $y$ and $z$ are the components of 3-D Cartesian coordinates. This paraboloid shape is chosen to make the spacial arrangement of the cells close to real biofilm while keeping the cell placement process mathematically simple. Within this 3-D biofilm region, we model the diffusivity according to $D_B/D_{aq} =0.4$, which is the mean relative diffusion \cite{biofilmdiff} where $D_B$ and $D_{aq}$ are the average molecular diffusion coefficients of the biofilm and pure water, respectively.
Further, to start the simulation at a stage where the biofilm is fully formed and the MC is already taking place, we filled the internal memory vector of each cell with the average molecular level at the initial TS. Each bacterial cell will use the initial signals from the internal memory and use its GRNN to process and update the feature vector for the next TS. Table \ref{table:Parameters} presents the parameter descriptions and values used for the simulation. As shown in Table \ref{table:Parameters}, the model runs for 150 TSs, generating data on a range of functions for the system. For instance, this model can produce data on feature vector of each cell, MC between cells, molecular consumption by cells, secretion to the environment, and nutrient accessibility of cells for each TS.
 \begin{figure}[t!]
\centering

\subfloat[\label{fig:Env20}]{
\includegraphics[trim={140 15 35 2},clip,width=0.42\linewidth]{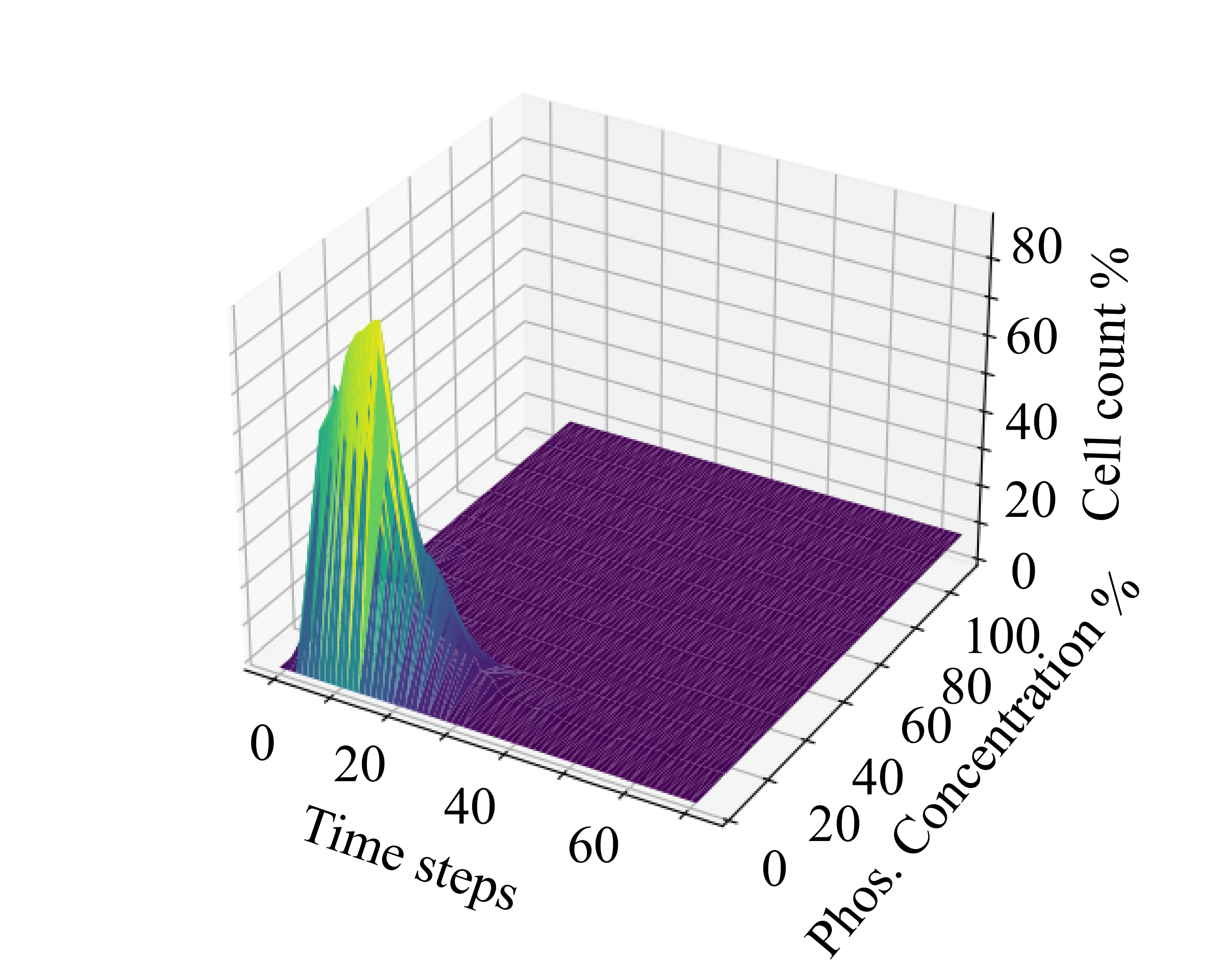}}
\subfloat[\label{fig:Env100}]{
\includegraphics[trim={120 15 50 2},clip,width=0.42\linewidth]{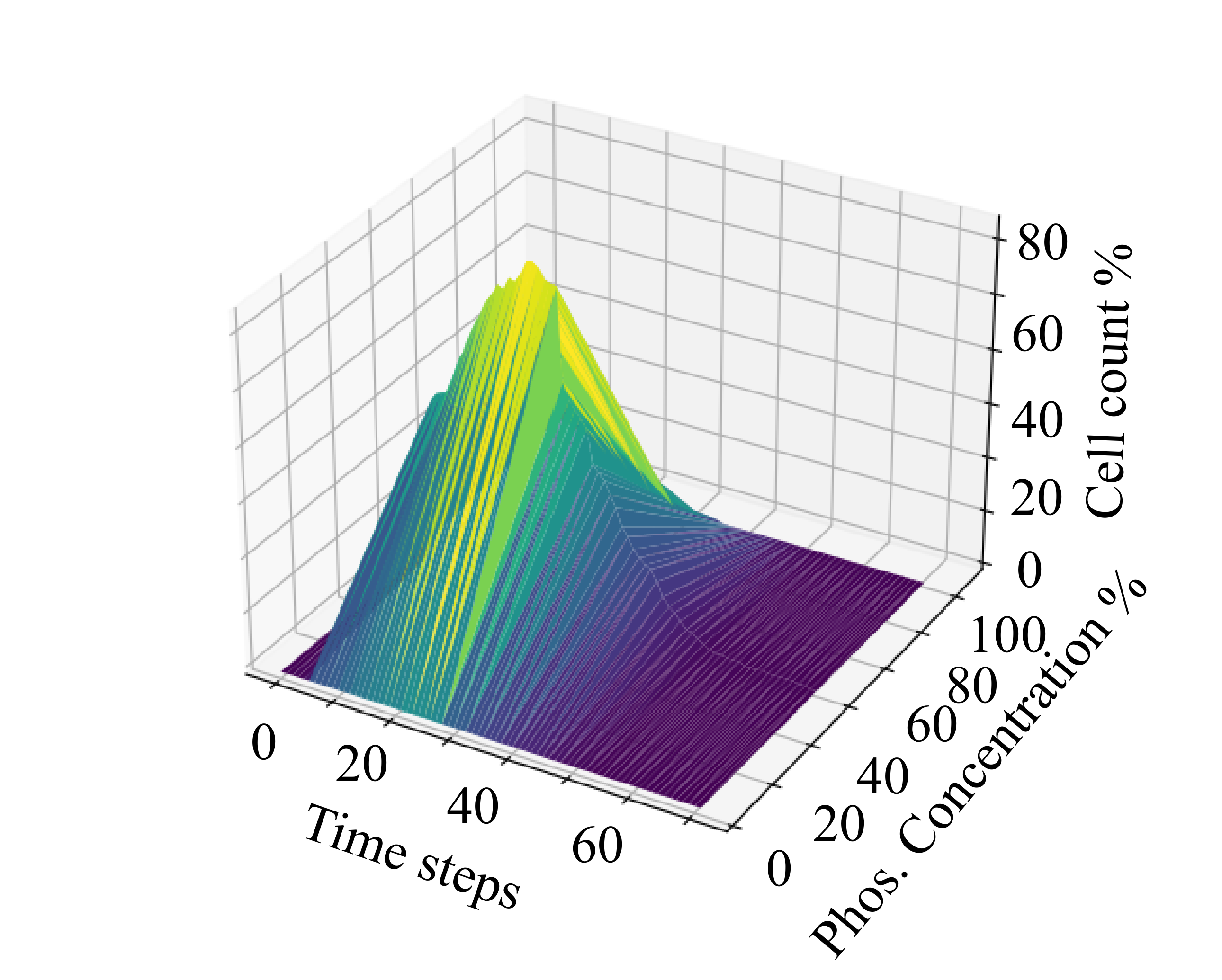}}
\caption{The nutrient accessibility variations of cells is expressed in two different environment conditions: a) low phosphate and b) high phosphate concentrations.\vspace{-1.5em}}
\label{fig:NutrientAccessibility}
\end{figure}

\begin{figure}[t!]
\centering

\subfloat[\label{fig:PyocyaninPOA1_LPHP}]{
\includegraphics[trim={11 0 10 0},clip,width=0.47\linewidth]{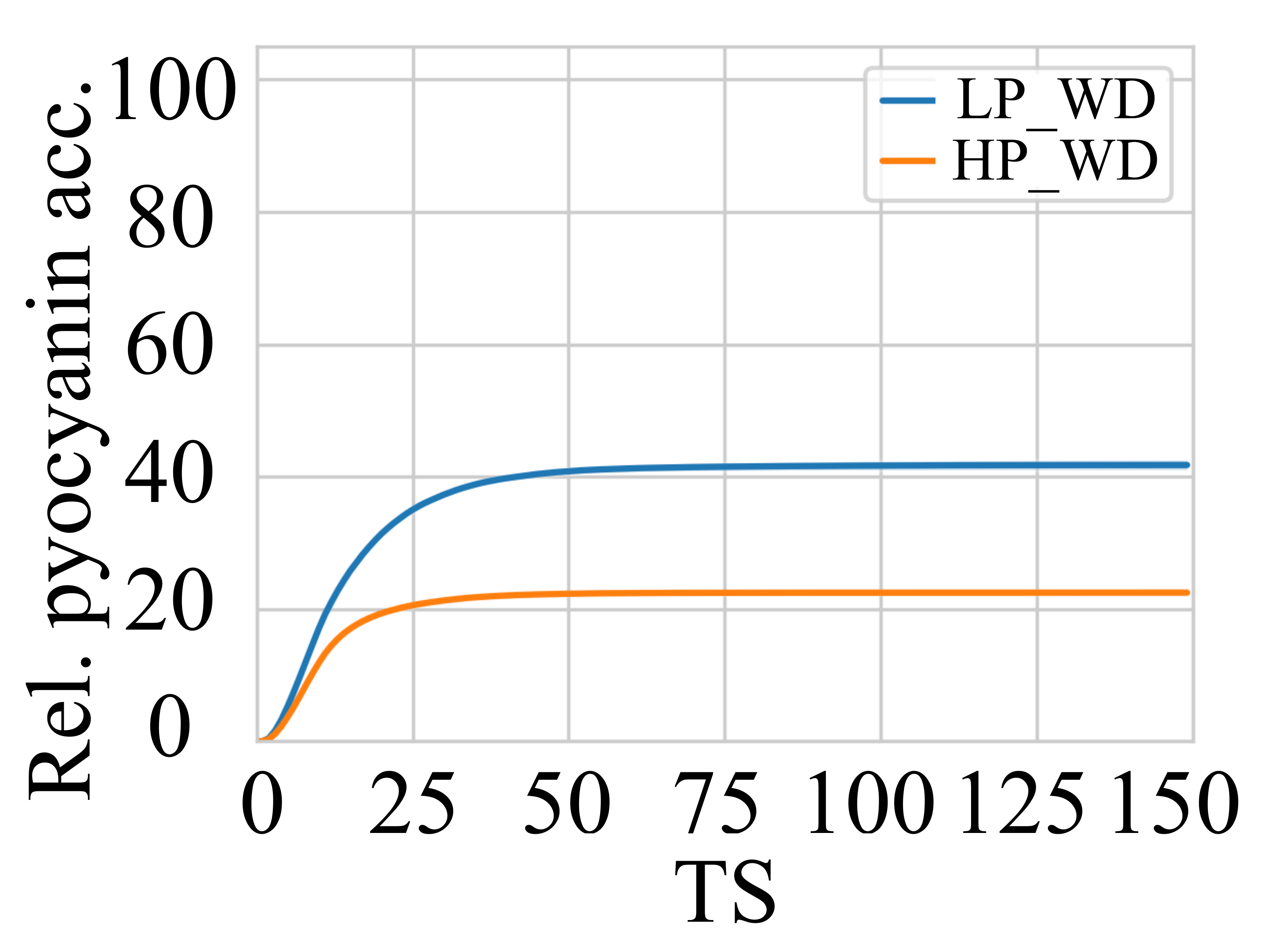}}
\subfloat[\label{fig:PyocyaninPOA1_LPHP_LASR}]{
\includegraphics[trim={11 0 10 0},clip,width=0.47\linewidth]{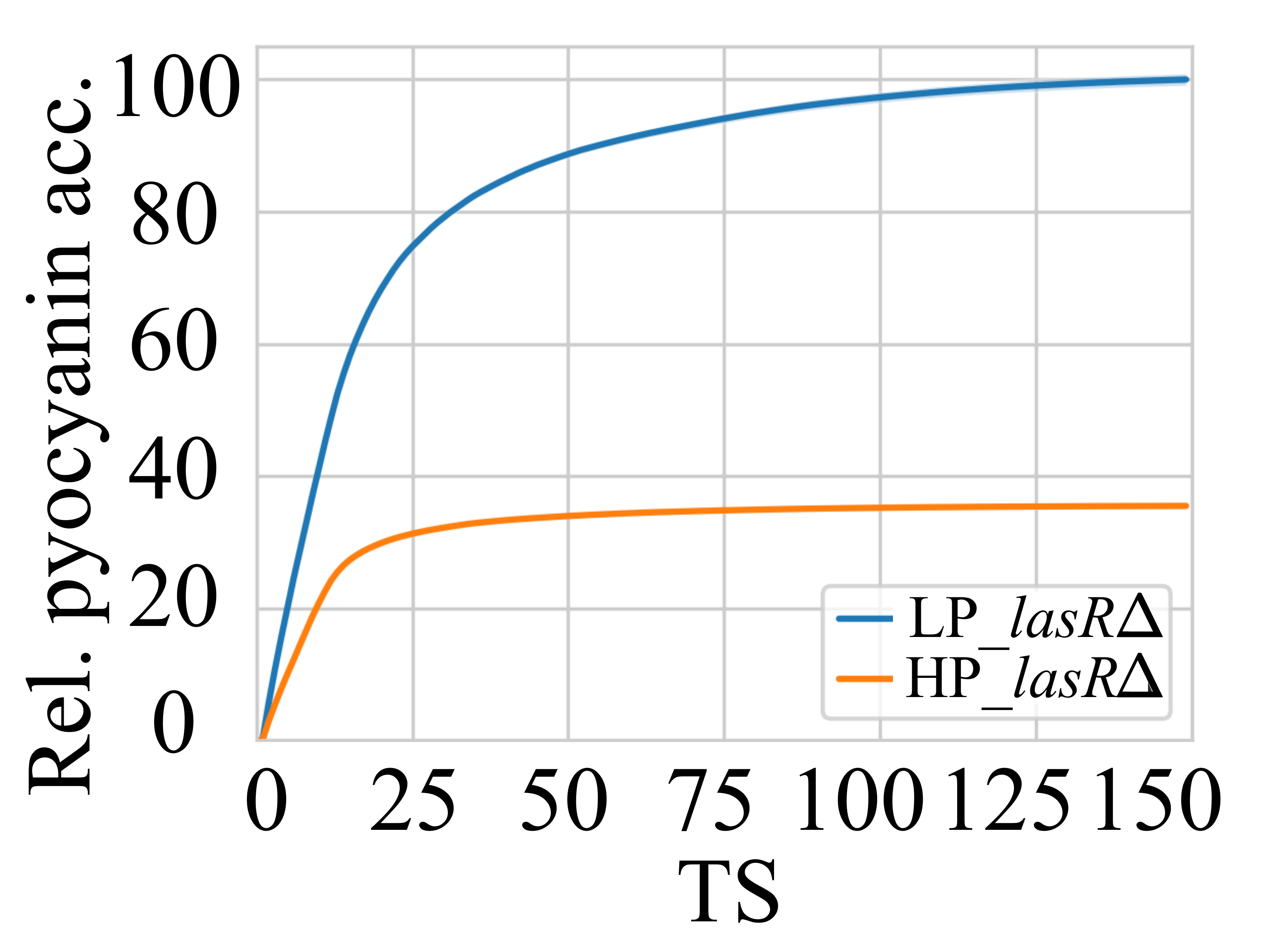}}

\subfloat[\label{fig:PyocyaninPOA1_LPHP_PHOB}]{
\includegraphics[trim={11 0 10 0},clip,width=0.47\linewidth]{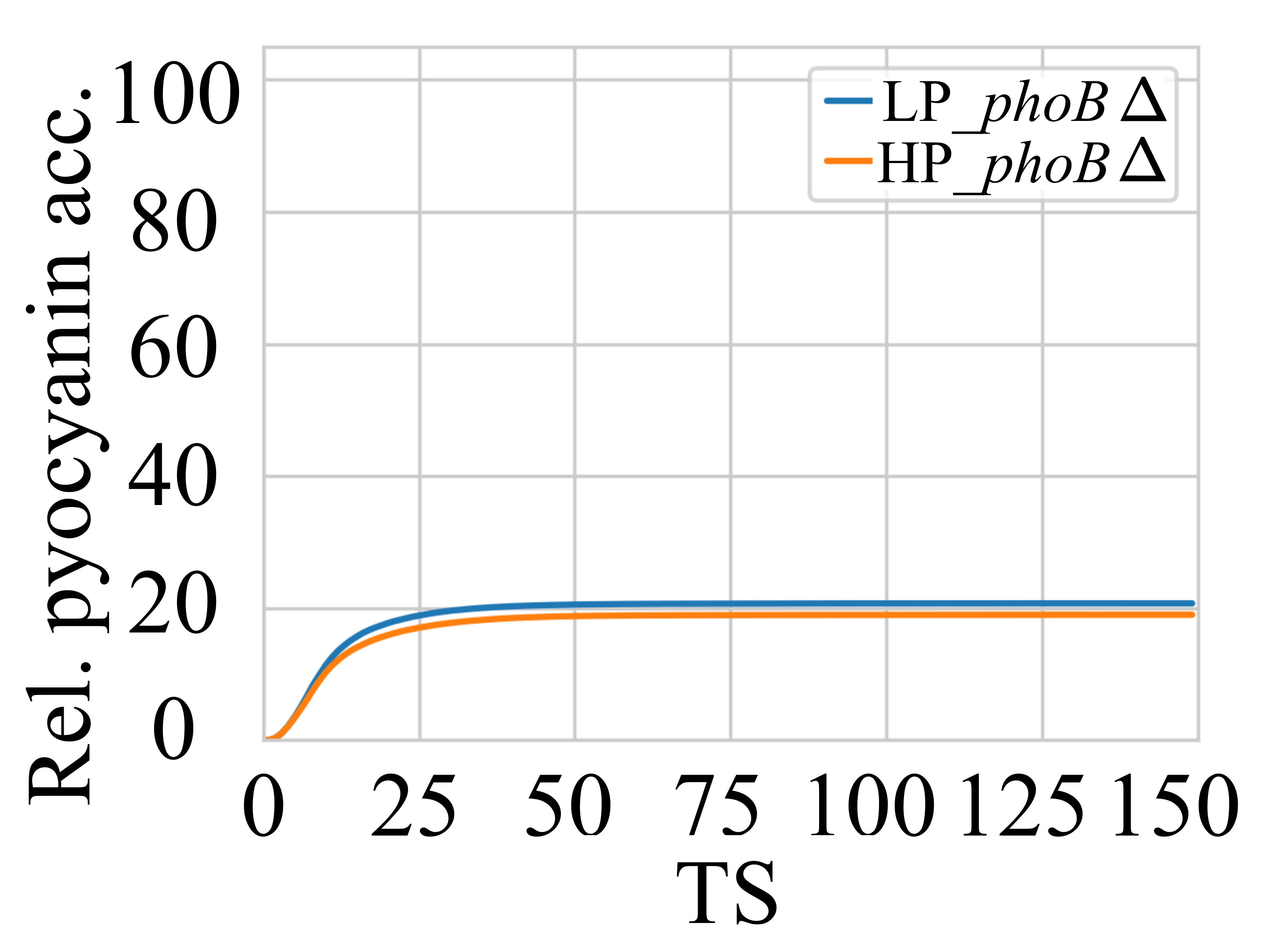}}
\subfloat[\label{fig:PyocyaninPOA1_LPHP_LASR_PHOB}]{
\includegraphics[trim={11 0 10 0},clip,width=0.48\linewidth]{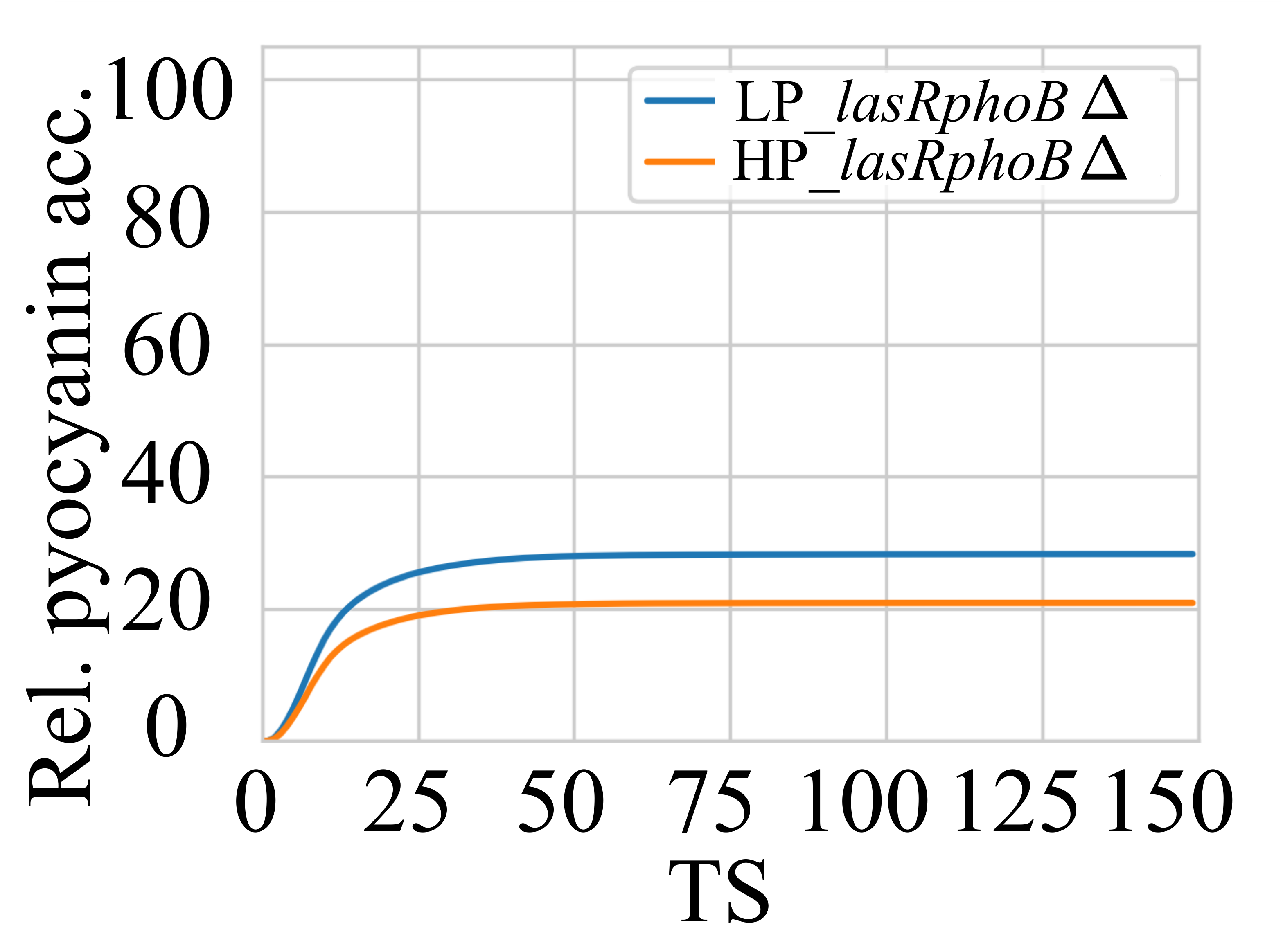}}
\caption{Relative Pyocyanin accumulation of four different biofilms of a) WD, b) lasR$\Delta$, c) phob$\Delta$ and d) lasR$\Delta$phob$\Delta$ in both low and high phosphate levels. \vspace{-1.5em}}
\label{fig:LP_HP_Pyocyanin}
\end{figure}

In order to prove that our GRNN computes similarly to the natural bacterial cell and collective behaviors of the cells are the same as the natural biofilm, we conduct a series of experiments. We explore the GRNN computation and biofilm activities under High Phosphate (HP) and Low Phosphate (LP) levels using eight experimental setups as follows, 1) wild-type bacteria (WD) in LP, 2) \textit{lasR} mutant (\textit{lasR$\Delta$}) in LP, 3) phoB mutant (\textit{phoB$\Delta$}) in LP, 4) \textit{lasR} \& PhoB double mutant (LasR$\Delta$PhoB$\Delta$) in LP, 5) WD in HP, 6) \textit{lasR$\Delta$} in HP, 7) \textit{PhoB$\Delta$} in HP and \textit{LasR$\Delta$PhoB$\Delta$} in HP. While the WD uses the full GRNN, \textit{lasR$\Delta$} is created by making the weight of the link between \textit{hn22} and \textit{lasR} as ``0''. Further, the GRNN of \textit{phoB$\Delta$} is created by making the weights of links from \textit{PhoB} to \textit{hn23} and \textit{PhoB} to \textit{pqsABCDE} also ``0''.\vspace{-1em}

\subsection{Model Validation}
\begin{figure}[t!]
    \centering
    \includegraphics[trim={0 0 0 0}, clip, width=0.8\linewidth]{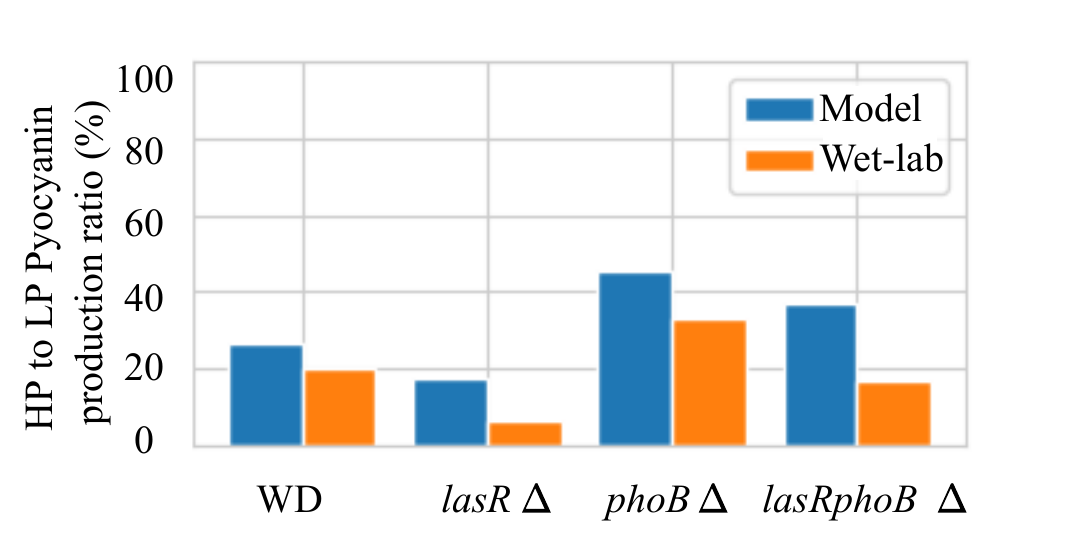}
    \caption{Evaluation of the model accuracy by comparing HP to LP pyocyanin production ratio with wet-lab data from \cite{meng2020molecular}. \vspace{-2.8em}}
    \label{fig:AccuracyComparison}
\end{figure}
First, we show the nutrient accessibility variation in the biofilm through Fig \ref{fig:NutrientAccessibility}. The cells in the biofilm core have less accessibility while the cells closer to the periphery have more access to nutrients due to variations in diffusion between the environment and the EPS. Fig \ref{fig:Env20} shows that when a low phosphate concentration is introduced to the environment, the direct access to the nutrient by the cells is limited. After the $TS=10$, around 60\% of cells have access to 20\% of the nutrient concentration. Further, Fig \ref{fig:Env100} shows that the increased nutrient introduction to the environment positively reflects on the accessibility. This accessibility plays a role mainly in the deviation of gene expression patterns resulting in phenotypic differentiations that is further analyzed in Section \ref{sec:GRNNComputingVariations}.
\begin{figure*}[t]
\centering
\subfloat[\label{fig:LasI}]{
\includegraphics[trim={11 5 10 50},clip,width=0.24\textwidth]{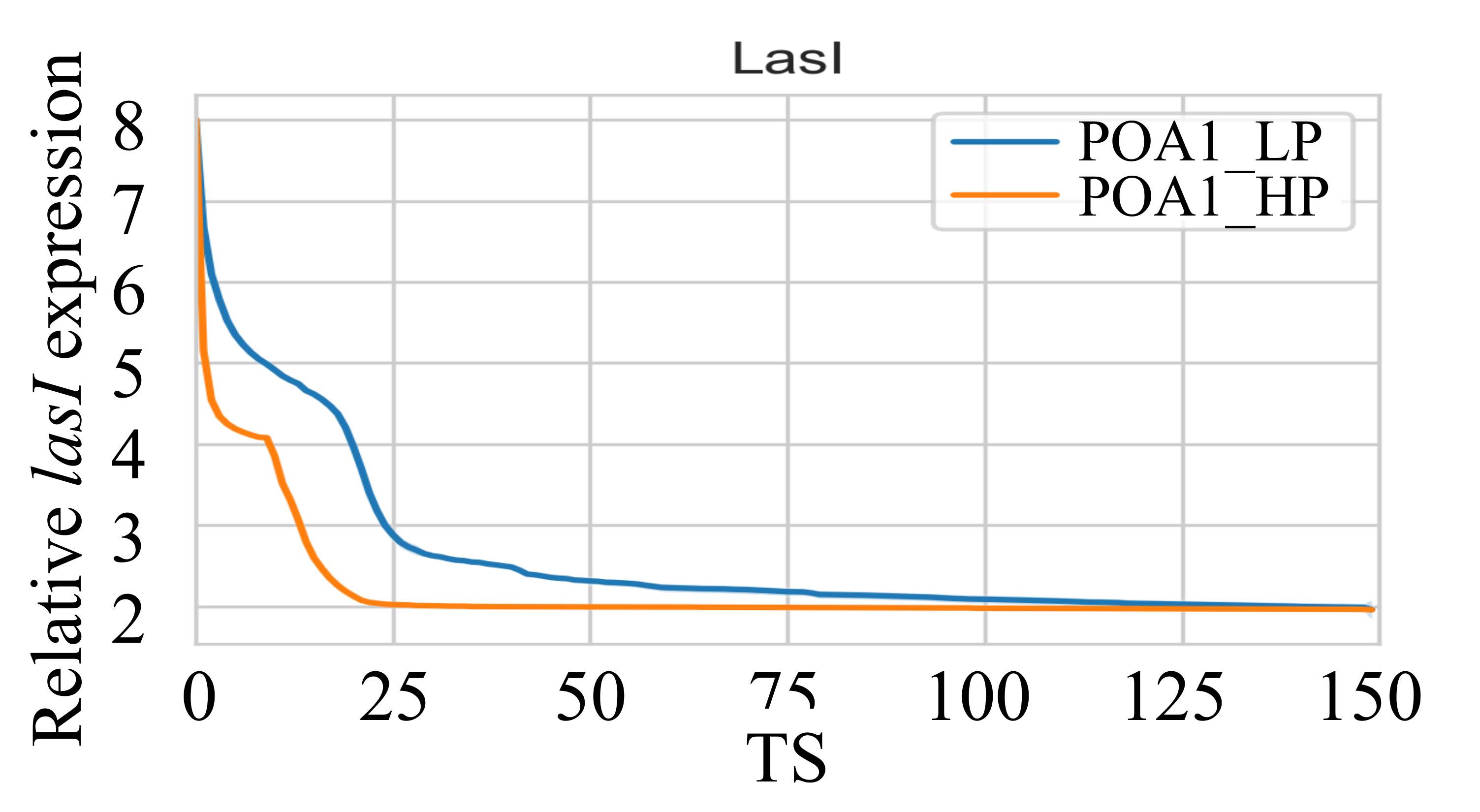}}
\subfloat[\label{fig:PqsA}]{
\includegraphics[trim={11 5 10 50},clip,width=0.24\textwidth]{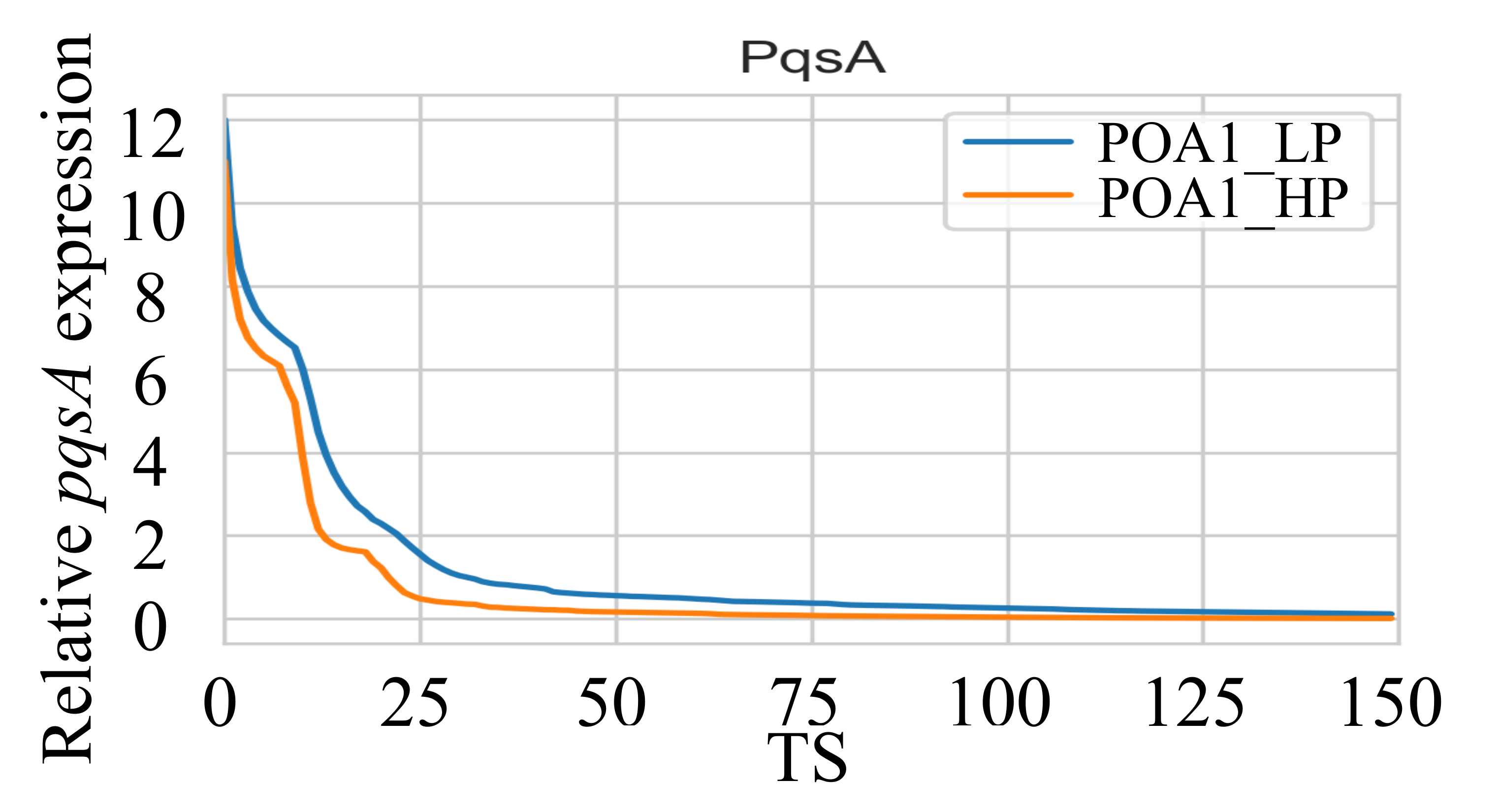}}
\subfloat[\label{fig:RhlR}]{
\includegraphics[trim={11 10 10 50},clip,width=0.24\textwidth]{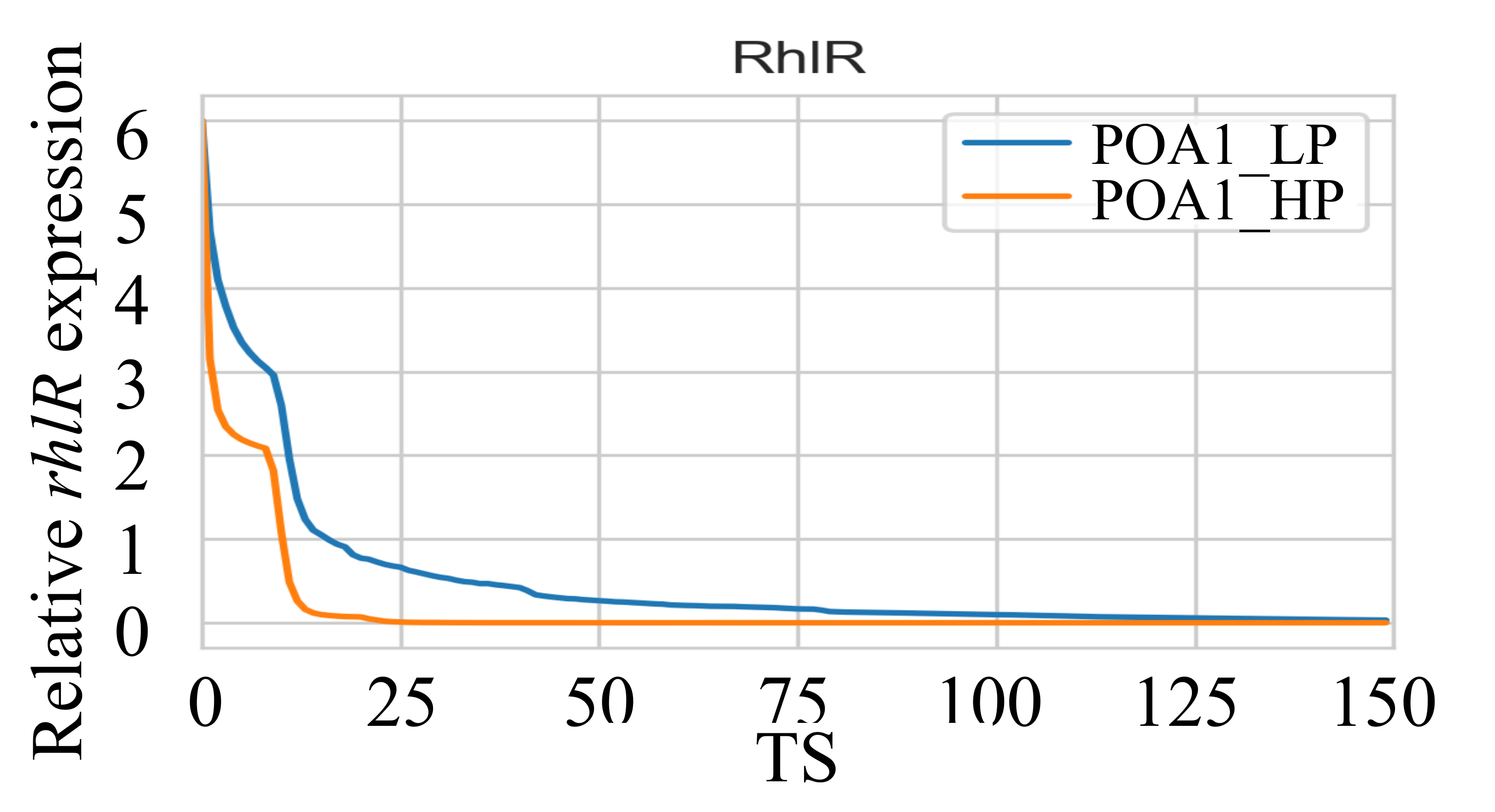}}
\subfloat[\label{fig:GEAcuuracy}]{
\includegraphics[trim={0 5 10 10},clip,width=0.26\textwidth]{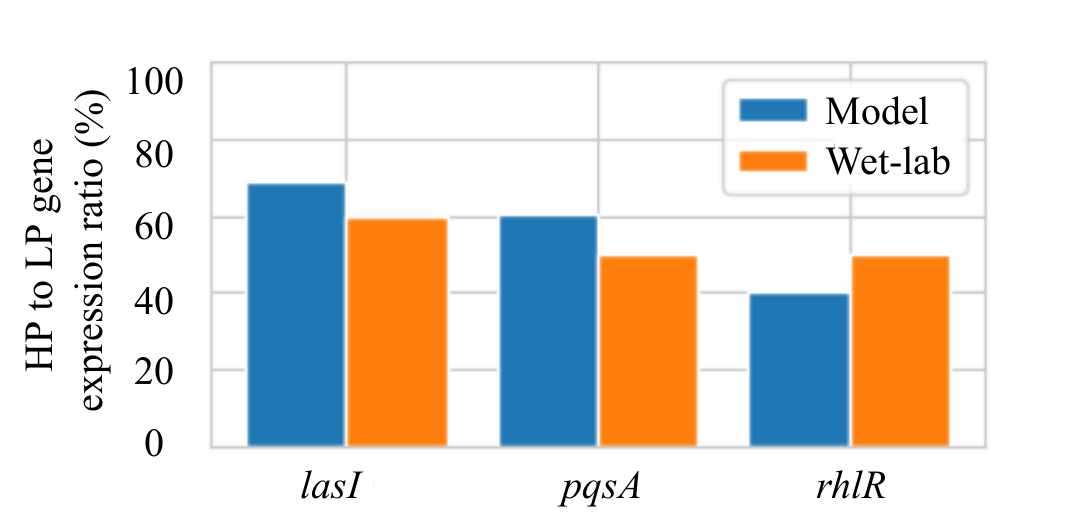}}
\caption{Expression levels of three different genes to that were used to prove the accuracy of the GRNN: a) \textit{lasI}, b) \textit{pqsA}, c) \textit{rhlR} expression levels in LP and HP and d) comparison between GRNN computing results and wet-lab data.}
\label{fig:Gene_expressions}
\end{figure*}

\begin{figure*}[t!]
\centering

\subfloat[\label{fig:NN_LP_OutBF}]{
\includegraphics[trim={0 9 0 0},clip,width=0.34\textwidth]{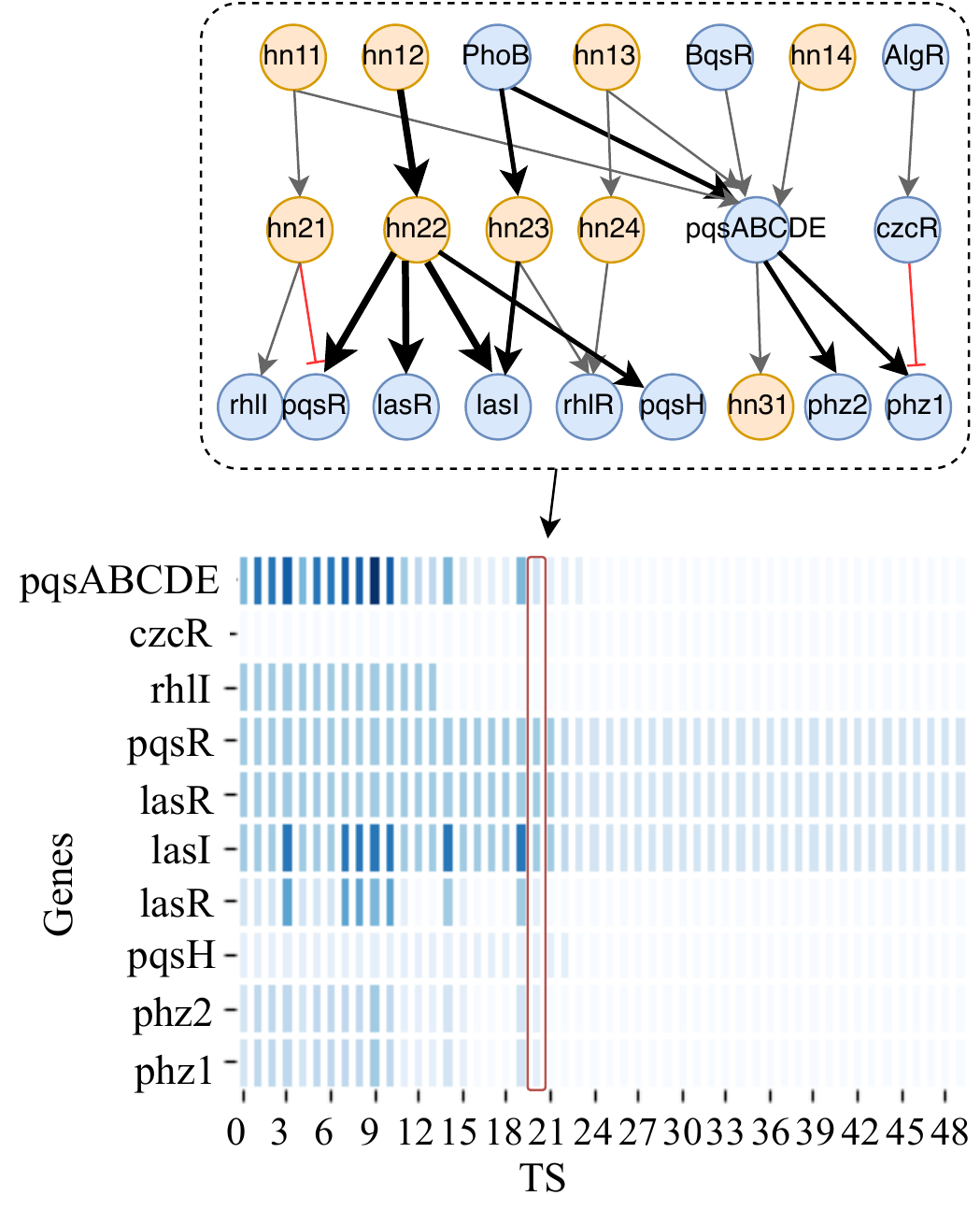}}
\subfloat[\label{fig:NN_LP_LASR_OutBF}]{
\includegraphics[trim={0 9 0 0},clip,width=0.27\textwidth]{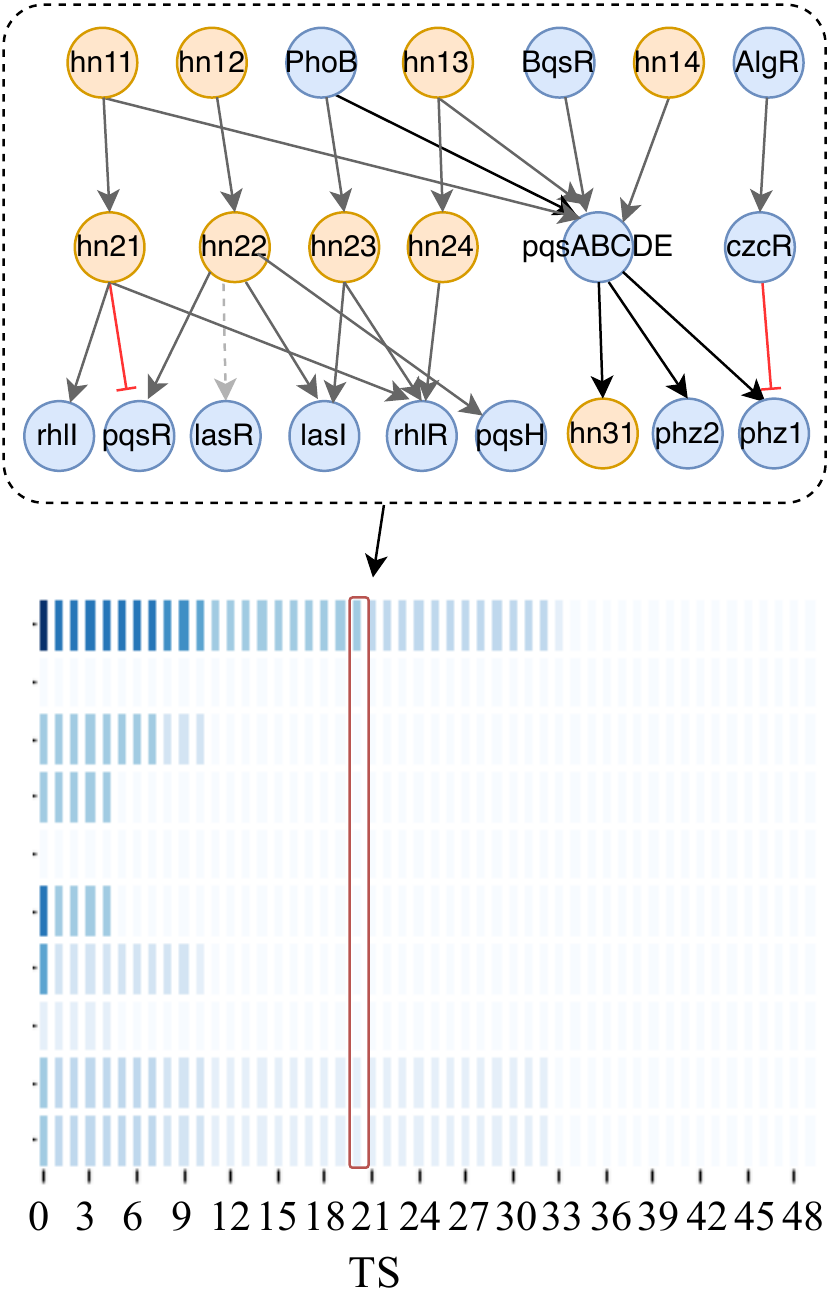}}
\subfloat[\label{fig:NN_LP_PHOB_OutBF}]{
\includegraphics[trim={0 9 0 0},clip,width=0.27\textwidth]{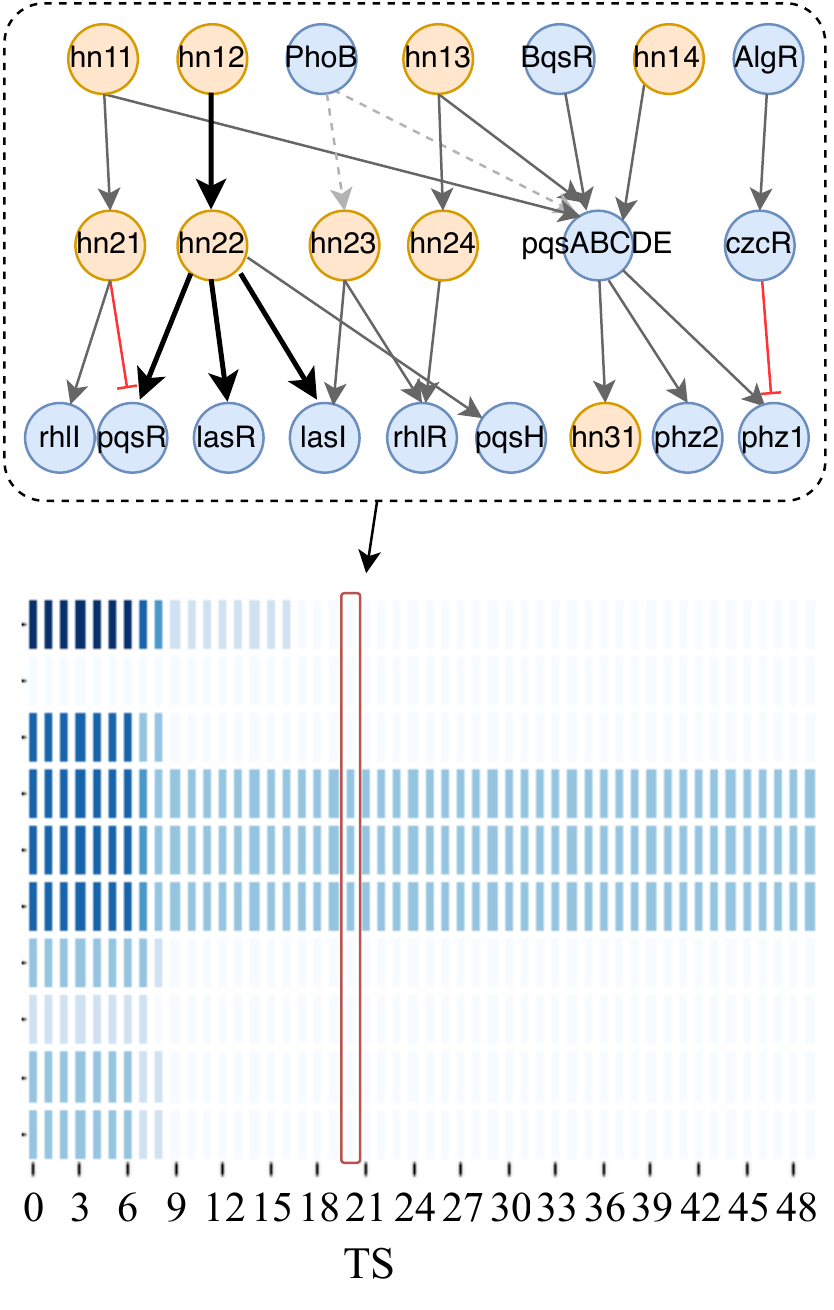}}
\caption{Gene expression and associated information flow variations in GRNNs of a) WD b) \textit{lasR$\Delta$} and c) phoB$\Delta$ in LP. \vspace{-1em}}
\label{fig:GEVsNN_LP}
\end{figure*}

\begin{figure*}[t!]
\centering

\subfloat[\label{fig:NN_HP_OutBF}]{
\includegraphics[trim={0 9 0 0},clip,width=0.345\textwidth]{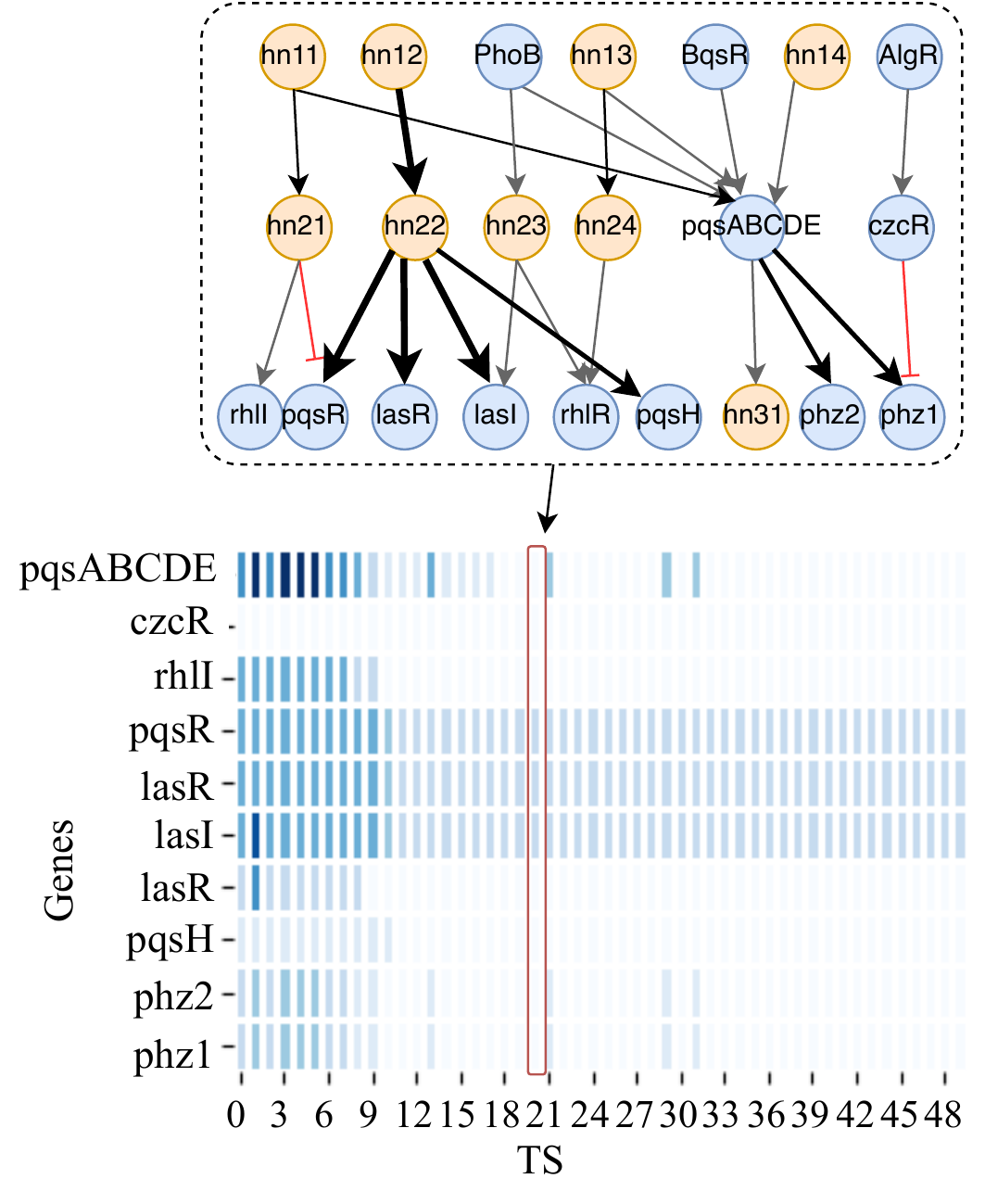}}
\subfloat[\label{fig:NN_HP_LASR_OutBF}]{
\includegraphics[trim={0 9 0 0},clip,width=0.28\textwidth]{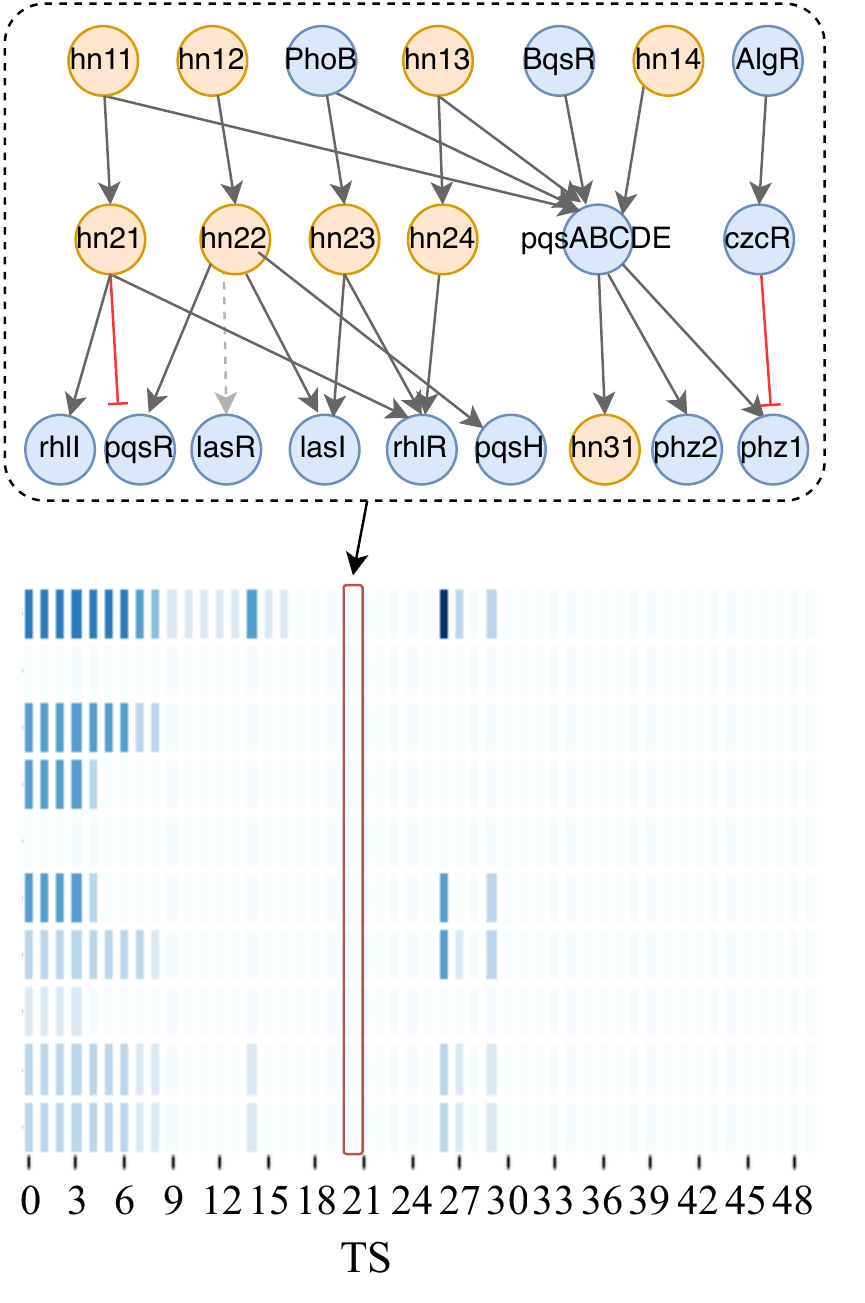}}
\subfloat[\label{fig:NN_HP_PHOB_OutBF}]{
\includegraphics[trim={0 9 0 0},clip,width=0.27\textwidth]{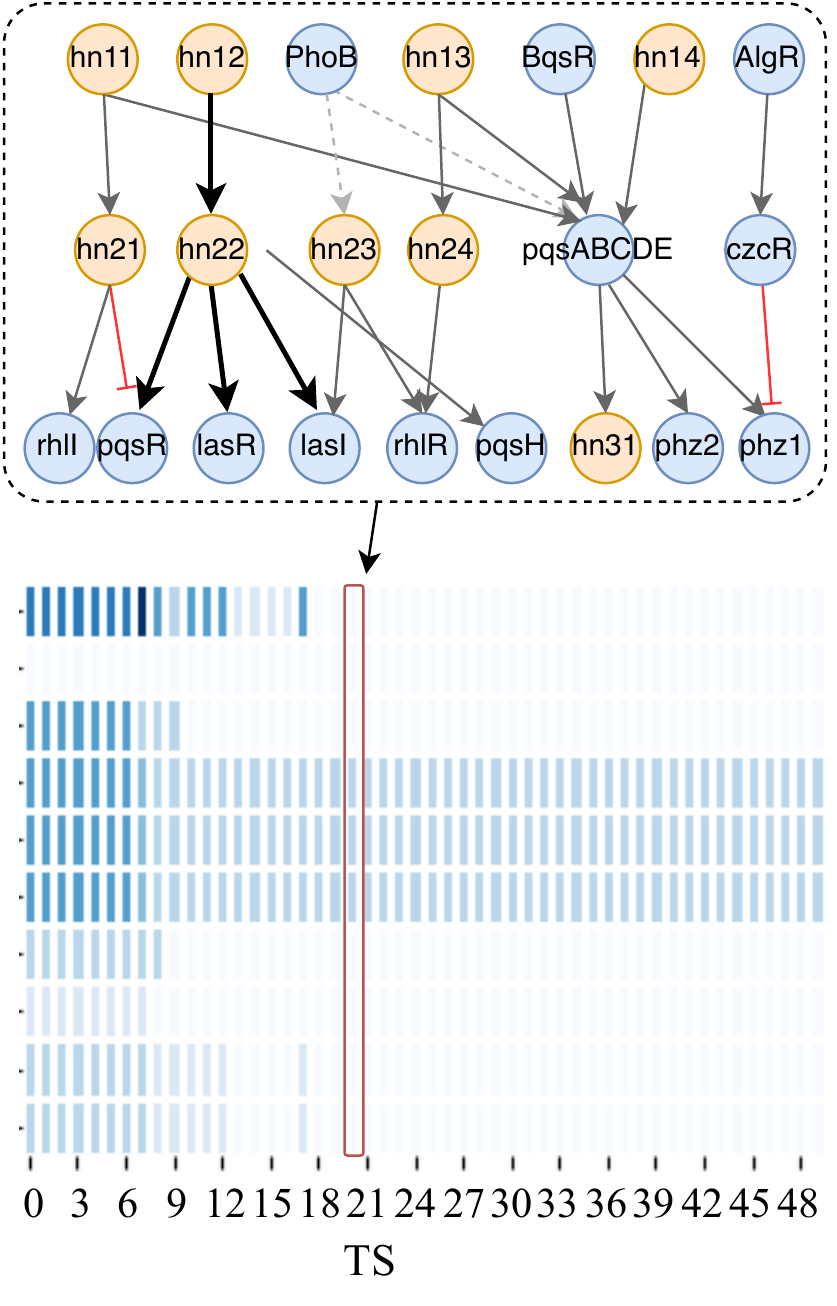}}

\caption{Gene expression and associated information flow variations in GRNNs of a) WD b) \textit{lasR$\Delta$} and c) phoB$\Delta$ in HP. \vspace{-1.5em}}
\label{fig:GEVsNN_HP}
\end{figure*}

\begin{figure*}[t!]
    \centering
    \includegraphics[trim={0 0 0 0}, clip, width=1\textwidth]{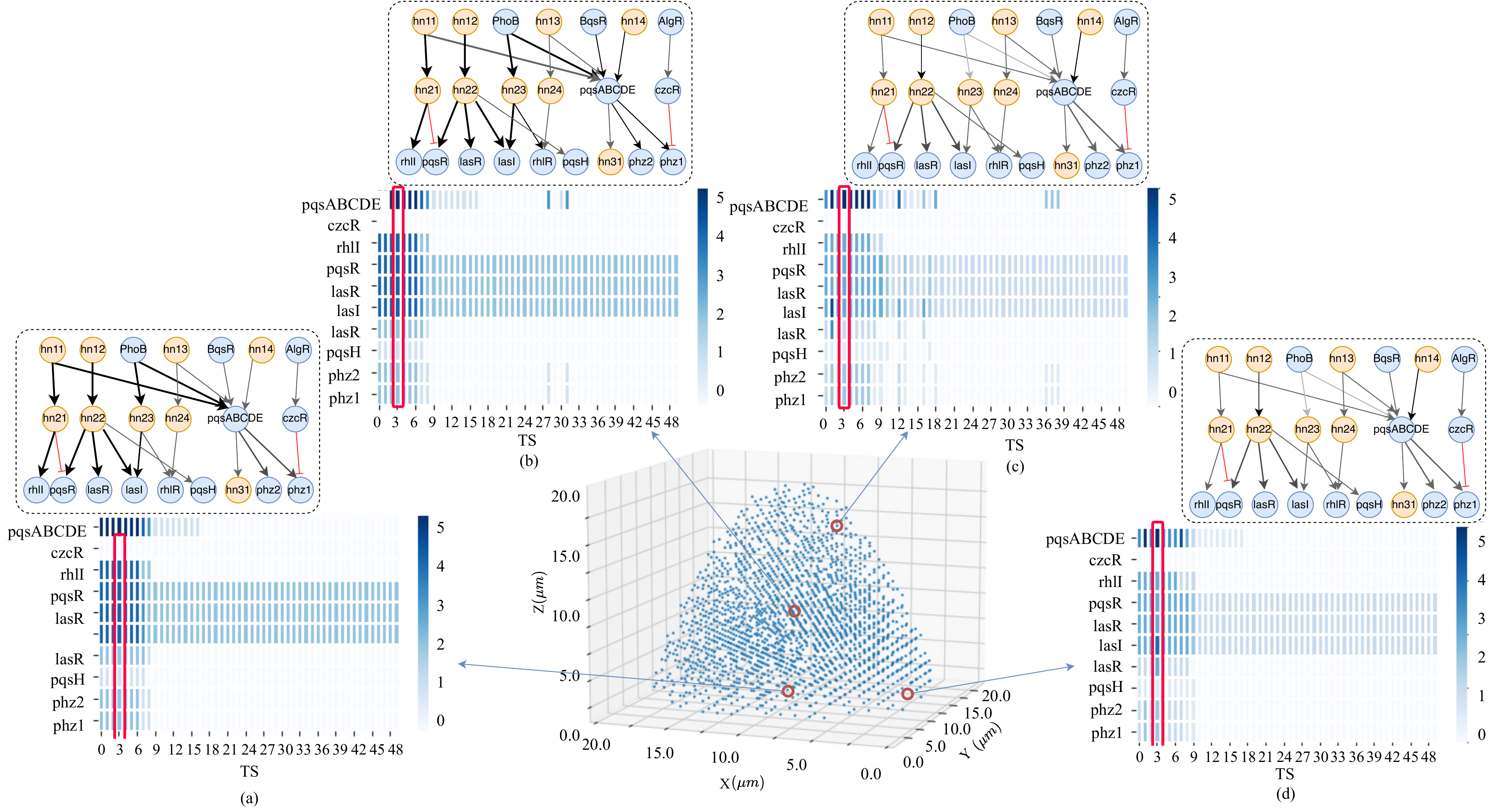}
    \caption{Illustration of GRNN information flow variations concerning the particular positions of cells within the biofilm. We selected four cells at a) [10, 10, 0] – close to the attached surface, b) [10, 10, 5]- close to the periphery, c) [7, 10, 13] – at the center and d) [3, 15, 0] – close to the attached surface and the periphery of the biofilm. \vspace{-1em}}
    \label{fig:DiffLocDiagram}
\end{figure*}
Comparing the predictions of molecular production through GRNN computing with the wet-lab experimental data from the literature, we are able to prove that the components of the GRN work similarly to a NN. Fig. \ref{fig:LP_HP_Pyocyanin} shows the pyocyanin accumulation variations of the environment in the eight setups mentioned earlier as results of decision-making of the GRNN. 
Production of pyocyanin of the WD \textit{P. aeruginosa} biofilms is high in LP, compared to the HP environments as shown in Fig: \ref{fig:PyocyaninPOA1_LPHP}. Further, the same pattern can be observed in the \textit{lasR$\Delta$} biofilms, but with a significantly increased pyocyanin production in LP as shown in Fig. \ref{fig:PyocyaninPOA1_LPHP_LASR}. The \textit{phob$\Delta$} and \textit{LasR}$\Delta$\textit{phob}$\Delta$ biofilms produce a reduced level of pyocyanin compared to WD and \textit{LasR}$\Delta$ that are shown in Fig. \ref{fig:PyocyaninPOA1_LPHP_PHOB} and Fig. \ref{fig:PyocyaninPOA1_LPHP_LASR_PHOB} respectively. 
We then present a comparison between GRNN prediction and wet-lab experimental data \cite{meng2020molecular} as ratios of HP to LP in Fig. \ref{fig:AccuracyComparison}. The differences between pyocyanin production through GRNN in HP and LP condition for all the four setups in Fig. \ref{fig:AccuracyComparison} are fairly close to the wet-lab data. In the WD setup, the difference between the GRNN model and wet-lab data only has around 5\% difference, while deviations around 10\%  can be observed in \textit{lasR$\Delta$} and \textit{phoB$\Delta$}. The most significant deviation around 20\% of pyocyanin production difference is visible in \textit{lasR$\Delta$phoB$\Delta$} that is caused by the lack of interaction from other gene expression pathways, as we only extracted a sub-network portion of the GRN. Therefore, these results prove that the extracted GRNN behaves similarly to the GRN dynamics.

\begin{figure*}[t!]
    \centering
    \includegraphics[trim={0 8 0 10}, clip, width=0.95\textwidth]{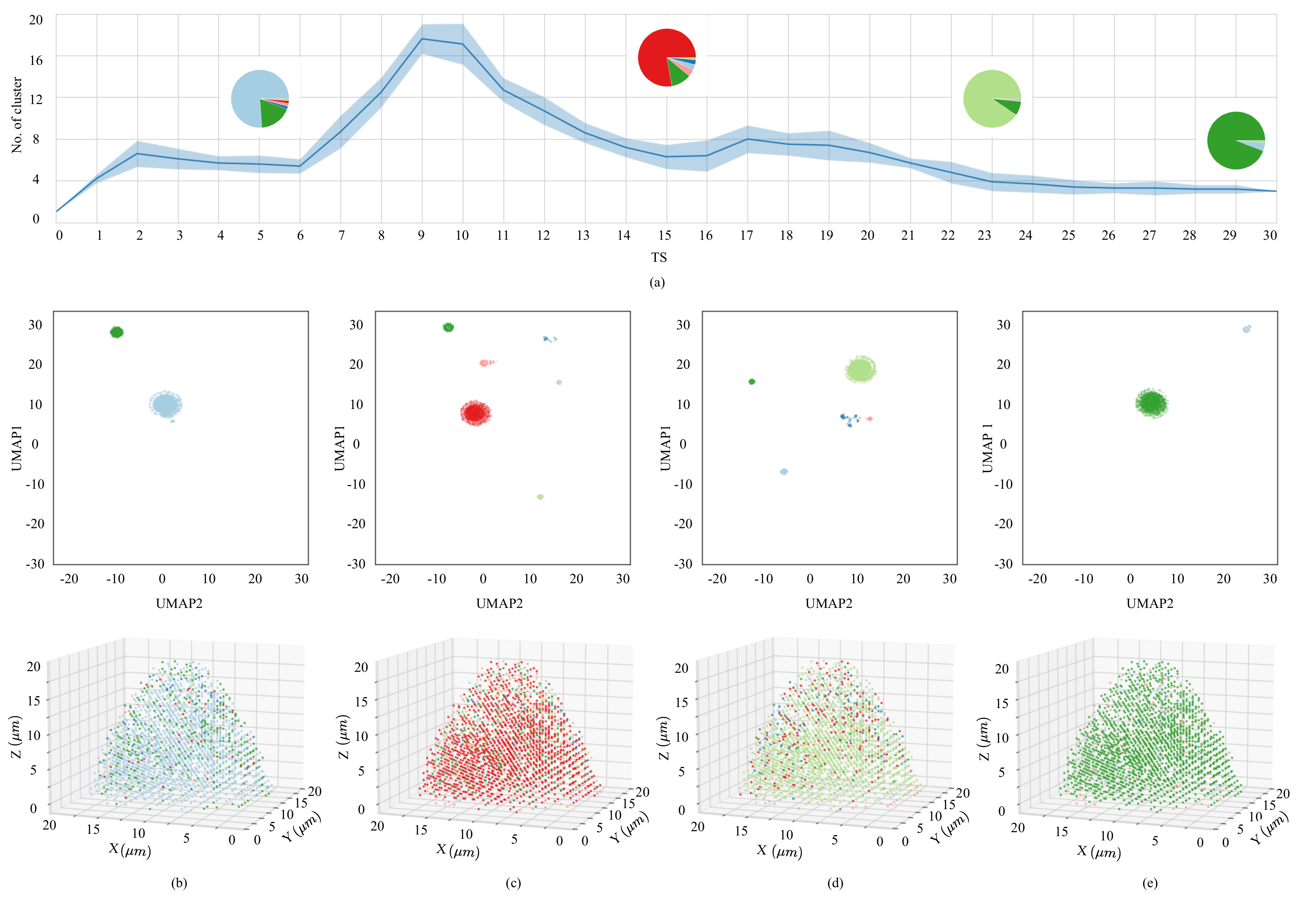}
    \caption{Illustration of GRNN-driven phenotypic cluster formation behaviors. a) shows the number of clusters (with their proportions via pie charts) for $TS<30$, b), c), d) and e) are pairs of the UMAP clustering based on gene expressions of cells and their locations in the biofilm at $TS=5$, $TS=15$, $TS=23$ and $TS=30$, respectively. \vspace{-1em}}
    \label{fig:ClustVsBFdrawioFull}
\end{figure*}

We further prove that the GRNN computing process performs similarly to the GRN by comparing the gene expression behaviors of the model with the wet-lab data \cite{meng2020molecular} as shown in Fig. \ref{fig:Gene_expressions}. First, we show the expression dynamics of genes \textit{lasI}, \textit{pqsA} and \textit{rhlR} of WD in LP in Fig. \ref{fig:LasI}, Fig. \ref{fig:PqsA} and Fig. \ref{fig:RhlR} respectively. All the figures depict that gene expression levels are higher in LP compared to HP until around $TS=100$. Beyond that point, relative gene expression levels are close to zero as the the environment run out of nutrients. Moreover, the differences in gene expression levels predicted by the GRNN computing for LP and HP are also compared with the wet-lab data in Fig. \ref{fig:GEAcuuracy}. In this comparison, it is evident that the predicted gene expression differences of all three genes are close to the wet-lab data with only around 10\% variation. The performance similarities between the GRNN and real cell activities once again prove that the GRN has underpinning NN-like behaviors. \vspace{-1em}

\subsection{Analysis of GRNN Computing}
\label{sec:GRNNComputingVariations}
Fig. \ref{fig:GEVsNN_LP} and Fig. \ref{fig:GEVsNN_HP} are used to show the diverse information flow of the GRNN that cause the variations in pyocyanin production in LP and HP conditions, respectively. Here, we use gene expression profiles extracted from one bacterial cell located at (7, 9, 2)$\mu$m in the Cartesian coordinates that is in the middle region of the biofilm with limited access to the nutrients. 
First, the gene expression variations of WD, \textit{lasR}$\Delta$, and \textit{phob$\Delta$} bacterial cells in LP (Fig. \ref{fig:GEVsNN_LP}) and HP (Fig. \ref{fig:GEVsNN_HP}) are shown for $TS<50$. Next, the information flow through the GRNN is illustrated above each expression profile at time $TS=20$, where the variations will be discussed. In Fig. \ref{fig:NN_LP_OutBF},  impact of the inputs 3OC-LasR and phosphate cause higher expression levels of the nodes \textbf{hn12} and \textit{phoB} in the input layer that cascade the nodes \textit{phZ1}, \textit{phZ2}, \textit{pqsR}, \textit{lasR}, 3OC, \textit{rhlR} and \textit{PqsH} in the output layer at $TS=20$. Fig. \ref{fig:NN_LP_LASR_OutBF} has significantly higher \textit{pqsA} operon expression levels compared to HP conditions (Fig. \ref{fig:NN_HP_LASR_OutBF}), reflecting higher pyocyanin production that can be seen in Fig. \ref{fig:PyocyaninPOA1_LPHP_LASR}. 
Nevertheless, the reduced gene expression levels, except \textit{pqsA} operon, of \textit{lasR$\Delta$} biofilm in both LP (Fig. \ref{fig:NN_LP_LASR_OutBF}) and HP (Fig. \ref{fig:NN_HP_LASR_OutBF}) conditions compared to the other setups emphasize that the inputs via inter-cellular MC significantly alter GRNN computing outputs.
In contrast, only a smaller gene expression difference can be observed between the two setups of \textit{phob$\Delta$} in LP (Fig. \ref{fig:NN_LP_PHOB_OutBF} and \textit{phob$\Delta$} in HP (Fig. \ref{fig:NN_HP_PHOB_OutBF}) resulting in minimized pyocyanin production differences as shown earlier in Fig. \ref{fig:PyocyaninPOA1_LPHP_PHOB}.

The GRNN model supports the  understanding of the gene expression variations due to factors such as nutrient accessibility, where in our case is a single species biofilm. Fig. \ref{fig:DiffLocDiagram} depicts the variability in the gene expression levels for four different locations of the biofilm at $TS=3$. 
Fig. \ref{fig:DiffLocDiagram}a and Fig. \ref{fig:DiffLocDiagram}b are the gene expression profiles and the signal flow through GRNN pairs of two cells located close to the attached surface and the center of the biofilm. The phosphate accessibility for these two locations is limited. Hence, edges from \textit{phob} have a higher information flow compared to the other two cells near the periphery of the biofilm, which can be observed in Fig. \ref{fig:DiffLocDiagram}c and Fig. \ref{fig:DiffLocDiagram}d. The microbes in the center (Fig. \ref{fig:DiffLocDiagram}a) and the bottom (Fig. \ref{fig:DiffLocDiagram}b) mainly have access to the inter-cellular MCs, while the other two bacteria have direct access to the extracellular phosphate. 

This GRNN produced data can further be used to understand the spatial and temporal dynamics of phenotypic clustering of gene expressions which is important in predicting and diagnosis of diseases\cite{dutta2019graph}. Fig. \ref{fig:ClustVsBFdrawioFull} shows the phenotypic variation of WD biofilm in LP. Fig. \ref{fig:ClustVsBFdrawioFull}a shows the number of cluster variations over the first 30 $TSs$ when the significant phenotypic changes of the biofilm is evident. At around $TS=9$ and $TS=10$, the bacterial cells have the most diverse expression patterns due to the highest extracellular nutrient penetration (can be seen in Fig. \ref{fig:Env20}) to the biofilm and inter-cellular communications. 
Here we use four TSs ($TS=5$ - Fig. \ref{fig:ClustVsBFdrawioFull}b, $TS=15$ - Fig. \ref{fig:ClustVsBFdrawioFull}c, $TS=23$ - Fig. \ref{fig:ClustVsBFdrawioFull}d and $TS=30$ - Fig. \ref{fig:ClustVsBFdrawioFull}e) to analyze this phenotypic differentiation. Each pair of Uniform Manifold Approximation and Projection (UMAP) plot and diagram of cell locations of each cluster explain how nutrient accessibility contribute to the phenotypic clustering. Although at $TS=5$ (Fig. \ref{fig:ClustVsBFdrawioFull}) the average number of clusters is over four, there are only two major clusters that can be observed with higher proportions, as shown in the pie chart. Among the two major clusters (blue and green) of Fig. \ref{fig:ClustVsBFdrawioFull}b, the bacteria in the blue cluster can mostly be found in the center of the biofilm, while the green cluster cells are close to the periphery. Fig. \ref{fig:ClustVsBFdrawioFull}c and Fig. \ref{fig:ClustVsBFdrawioFull}d have more clusters as the nutrient accessibility among cells is high. In contrast, due to the lack of nutrients in the biofilm, a limited number of clusters can be seen in the biofilm after around $TS=30$, which can be observed Fig. \ref{fig:ClustVsBFdrawioFull}e.\vspace{-0.5em}


\section{Conclusion}
\label{sec:Conclusion}

The past literature has captured the non-linear signal computing mechanisms of Bacterial GRNs, suggesting underpinning NN behaviors. This study extracts a GRNN with summarized multi-omics gene expression regulation mechanisms as weights that can further analyze gene expression dynamics, design predictive models, or even conduct \textit{in-vivo} computational tasks. 
We used \textit{P. aeruginosa} single species biofilm as a use case and extracted relevant gene expression data from databases such as RegulomePA and transcriptomic data from databases including GEO. Due to the complexity of the GRN and expression dynamics, we only considered a smaller sub-network of the GRN as a GRNN that is associated with QS, iron and phosphate inputs, and pyocyanin production. Considering this GRNN, we modeled the computation process that drives cellular decision-making mechanism. As bacteria live in ecosystems in general where intra-cellular communication play a significant role in cellular activities, an \textit{in-silico} biofilm is modeled using GNN to further analyze the biofilm-wide decision-making.
A comparison between the GRNN generated data and the transcriptomic data from the literature exhibits that the GRN behaves similarly to a NN. Hence, this model can explore the causal relationships between gene regulation and cellular activities, predict the future behaviors of the biofilm as well as conduct bio-hybrid computing tasks. Further, in the GRNN extraction phase, we were able to identify the possibility of modeling more network structures with various number of input nodes, hidden layers, and output nodes. In addition, GRN components including auto regulated genes and bidirectional intergenic interactions hints the possibility of extracting more sophisticated types of GRNNs such as Recurrent NN and Residual NN in the future.
The idea of extracting sub-networks as NNs can lead to more intriguing intra-cellular distributed computing. Further, this model can be extended to multi-species ecosystems for more advanced predictive models as well as  distributed computing architectures combining various NNs. \vspace{-1em}

\ifCLASSOPTIONcaptionsoff
  \newpage
\fi

\bibliographystyle{ieeetr}
\bibliography{sample}

\end{document}